\documentclass[conference]{IEEEtran}
\IEEEoverridecommandlockouts
\usepackage{amsmath,amssymb,amsfonts}
\usepackage{algorithmic}
\usepackage{graphicx}
\usepackage{textcomp}

\usepackage[backend=biber,style=ieee,natbib=true]{biblatex} 
\usepackage{subfigure}
\usepackage[table,xcdraw]{xcolor}
\usepackage{url}
\usepackage{balance}
\usepackage[brazil]{babel}

\usepackage[normalem]{ulem}
\usepackage{enumitem} 
\usepackage{subfigure} 
\usepackage[format=hang,font=footnotesize,justification=justified,figurename=Fig.,labelfont=bf]{caption}
\usepackage[utf8]{inputenc}

\def\BibTeX{{\rm B\kern-.05em{\sc i\kern-.025em b}\kern-.08em
    T\kern-.1667em\lower.7ex\hbox{E}\kern-.125emX}}

\begin{document}

\title{De Quem é o Jogo? Disputas Narrativas no Fandom de World of Warcraft}

\author{\IEEEauthorblockN{Clara Andrade Pimentel}
\IEEEauthorblockA{\textit{Depto. de Comunicação Social} \\
\textit{Universidade Federal de Minas Gerais}\\
Belo Horizonte, Brasil\\
clarapimentel@ufmg.br}
\and
\IEEEauthorblockN{Joana Ziller}
\IEEEauthorblockA{\textit{Depto. de Comunicação Social} \\
\textit{Universidade Federal de Minas Gerais}\\
Belo Horizonte, Brasil\\
joana.ziller@gmail.com}
\and
\IEEEauthorblockN{Philipe Melo}
\IEEEauthorblockA{\textit{Depto. de Ciência da Computação} \\
\textit{Universidade Federal de Minas Gerais}\\
Belo Horizonte, Brasil\\
philipe@dcc.ufmg.br}
}

\maketitle

\vspace{-3.5cm}

\begin{abstract}
Cada vez mais, os jogos digitais fazem parte de uma cibercultura engendrada pelas plataformas digitais. Pensando nisto, abordamos neste trabalho
algumas considerações a respeito dos jogadores de World of Warcraft como fãs e produtores de conteúdo e as disputas narrativas que emergem sobre o jogo em plataformas de publicação de trabalhos de fãs (Archive of Our Own e DeviantArt). Analisamos um vasto conjunto de \textit{fanfics} e \textit{fanarts} coletadas nestas plataformas, mostrando uma textualidade que envolve não somente o jogo digital, mas toda uma rede de produção dos fãs que se expande para além do ato de jogar. Nossas observações evidenciam que, apesar da percepção popular de que o fandom de World of Warcraft é majoritariamente masculino e heteronormativo, mulheres e pessoas LGBTQI+ são um grande público participativo e produzem muito conteúdo, especialmente no universo de \textit{fanfics}. As obras criadas também são bastante marcadas por narrativas de corpos e sexualidades dissidentes. Entretanto, apesar da presença desses sujeitos e narrativas no fandom, esse conteúdo é invisibilizado no DeviantArt, que privilegia artistas homens e \textit{fanarts} heteronormativas de caráter comercial.
\end{abstract}

\begin{IEEEkeywords}
World of Warcraft, Estudos de Fãs, Fanart, Fanfic, Gênero, Plataformas, DeviantArt, Archive of Our Own
\end{IEEEkeywords}

\section{Introdução}

A ideia de uma ``cultura dos jogos digitais'' ou então uma ``subcultura de jogos'' é delimitada comumente pela academia pela suposição de uma identidade comum compartilhada pelos jogadores (a identidade ``\textit{gamer}'') e práticas sociais constituídas por esses sujeitos em ambientes virtuais. Desse modo, pesquisadores desenvolveram o hábito de falar sobre uma cultura dos jogos trazendo uma visão interna, como muitos deles se auto-identificam como \textit{gamers}. Entretanto, os jogos são também constituídos como uma subcultura através da retratação midiática de quem são os \textit{gamers} e quais seus costumes. Assim, a cultura dos jogos foi sendo historicamente situada como predominantemente masculina através dos estudos críticos à indústria e público, das propagandas dos jogos com garotos jogando na sala de estar e manchetes jornalísticas ilustradas por fotografias de equipes masculinas de e-Sports ou que criticam a cultura masculina tóxica dos jogos~\cite{shaw2010videogameculture}.

Mulheres, pessoas LGBTQI+ e pessoas não-brancas foram, consequentemente, delegadas a um espaço de marginalização ao tratarmos dos jogos. O que jogam e como jogam não são considerados constituintes de uma cultura tradicional dos jogos. Além disso, a proficiência técnica necessária para o desenvolvimento é tradicionalmente atribuída aos homens brancos, fortalecendo a noção de uma identidade \textit{gamer} essencialmente masculina. Todavia, a realidade é mais complexa do que isso: os grupos marginalizados pelos discursos hegemônicos participam de comunidades de jogos, produzem jogos e também podem se reconhecer como \textit{gamers}. Isso demanda de nós, pesquisadoras e pesquisadores, uma leitura crítica da indústria e dos trabalhos acerca de uma suposta cultura dos jogos digitais.

Sendo assim, nesse artigo tomamos como ponto de partida o pressuposto de que os jogos não são, necessariamente, uma forma cultural distinta da cultura dominante. Essa visão é muitas vezes reiterada, mesmo que acidentalmente, por trabalhos que possuem foco no jogar, na identidade de jogadores e no conteúdo do jogo. Essa especificidade pode encapsular os efeitos do jogo na cultura e na vida das pessoas, muitas vezes os resumindo à experiência durante o ato de jogar. Seguindo esse raciocínio, buscamos aqui abordar o jogo como parte inerente de um arranjo cultural mais amplo, inserido em um contexto digital. Esta visão é fundamental, pois é percebido que a cibercultura, da qual os jogos fazem parte, em geral, é generificada como masculina e centrada no ocidente, através da exclusão de vozes femininas e periféricas. 


A partir disso, decidimos estudar como os fãs de \textit{World of Warcraft} (WoW) (\textit{Blizzard Entertainment}, 2004), um dos jogos mais populares das últimas décadas, relacionam-se com o jogo e suas histórias através do mapeamento de suas produções nas plataformas Archive of Our Own (AO3)\footnote{Archive of Our Own. Disponível em \url{https://archiveofourown.org/}.} e DeviantArt (dA)\footnote{DeviantArt. Disponível em \url{https://www.deviantart.com/}.}. A primeira é um importante espaço para publicação de \textit{fanfics} (ficções escritas de fãs) e a segunda é popularmente utilizada para postagem de \textit{fanarts} (artes de fãs). Procuramos chamar atenção a uma textualidade que envolve não apenas os jogos digitais, mas também os jogadores, produtores de conteúdo, diferentes mídias e textos. Para isso, utilizamos da coleta das produções de fãs e realizamos uma análise, com ênfase no gênero, de seus metadados, explorando aspectos dos jogos digitais que muitas vezes tornam-se invisibilizados pela arquitetura das redes sociais e discursos hegemônicos da mídia tradicional.

\section{Gamers e fãs}

A centralização dos enunciados produzidos sobre jogos ao redor dos homens e a desvalorização de outros corpos é comum mesmo além dos jogos digitais, fazendo-se presente em diferentes espaços de fãs e na produção cultural que os sustentam \cite{Gray2018}. Os fãs de cultura pop, por anos, foram tratados como pessoas patológicas e estereotipados pela mídia como nerds obcecados, com frustrações sexuais e desmasculinizados~\cite{jenkins1992textual}. Os jogadores, em particular, foram alvo de diversos discursos sobre como são indivíduos antissociais e o ato de jogar é constantemente associado a problemas de saúde e comportamentos obsessivos~\cite{shaw2010videogameculture}. Contudo, nos anos 2000, o fã masculino de cultura pop passou por uma reformulação através do reposicionamento do nerd como uma demografia de mercado \cite{christinequail_2009}.

Após esse reposicionamento, os comportamentos divergentes e patológicos permaneceram, mas reimaginados como espécies de apetrechos românticos do herói, aquilo que demonstra sua capacidade de amor e devoção intensos. Logo, essas características antes vistas como indesejadas, são colocadas como elementos que devem ser aceitos pela parceira em um relacionamento heteronormativo e que enfatizam a paixão latente dos homens. Essa reconfiguração do \textit{fanboy} consegue reintegrar os fãs homens numa espécie nova de masculinidade hegemônica~\cite{Scott2019}. Ainda assim, a imagem do fã não melhorou tanto para as mulheres. As fãs femininas continuam a ser representadas na mídia de maneira patológica, como mulheres que não conseguem se engajar em relacionamentos afetivos satisfatórios, ou que não valorizam suas carreiras~\cite{Scott2019}. É ingênuo pensarmos que não existem fãs mulheres, que elas não se engajam com os produtos culturais, ou que não dominam o conhecimento acerca das obras. 

Quando falamos de fãs falamos de pessoas que se engajam em práticas em torno de determinados produtos culturais. Nos interessa aqui os fãs de jogos, sujeitos que, além de jogar, assistem a \textit{streamings}, torcem por seus times de e-Sports favoritos, comentam em redes sociais, usam produtos de suas franquias prediletas e, sobretudo, criam seu próprio conteúdo para que outros possam ver e compartilhar. Ou seja, os sujeitos que produzem textos e práticas que constituem uma cultura de jogos para além do ato de jogar.
Segundo \citet{jenkins1992textual}, o comportamento do fã é marcado por seu vínculo ativo com o conteúdo através de atividades criativas e comunitárias -- como a desenho de \textit{fanarts}, remixes de vídeos e áudio, escrita, organização de eventos, etc. O conjunto dessas atividades, as relações entre os indivíduos e os produtos culturais criados conformam o que entendemos como fandom, termo popularmente atribuído às comunidades de fãs, associado aos estudos culturais norte-americanos.

Ao definirem o que é um fandom, \citet{Brough2011} tomam como fator de coesão para o grupo uma identidade coletiva ou subcultural formada em torno de interesses similares. Essa unidade de fãs conseguiria resistir através de suas produções a um status quo, ou mesmo apresentar ideias progressistas, como abordado pelos estudos de fãs inicialmente~\cite{Stanfill2020}. Porém, é importante apontarmos que essa é uma visão romantizada da cultura participativa, já que nem sempre participação significa resistência~\cite{Jenkins2015}. Essa perspectiva idealizada pode legitimar ``as assimetrias de poder entre produtores e consumidores de cultura''~\cite{Garson2019} ao não reconhecer as nuances das relações entre os próprios fãs e instituições. 


As  fronteiras do fandom são constantemente vigiadas por fãs que avaliam a legitimidade do conhecimento do outro, validando se uma pessoa é uma fã ``verdadeira'' ou não. \citet{JohnsonFantagonism} explica como a impressão de um fandom unificado e homogêneo só é possível após uma disputa interna de discursos, em que uma visão emerge como hegemônica, mesmo que temporariamente; \citet{Stanfill2020} vai além, apontando como o fandom, inclusive, pode ser um espaço reacionário ao invés de progressista. Um fandom, portanto, comporta uma competição antagonística entre discursos de diferentes grupos, em que há um esforço contínuo pela legitimação e deslegitimação de cada narrativa e, por consequência, a exclusão de certos sujeitos desses espaços. Essas distinções de conteúdos são um reflexo pela busca das pessoas por estabilidade e coerência de suas próprias identidades, espelhando suas experiências sociais e expectativas sobre o produto cultural favorito~\cite{fisk1994culturaleconomy}.

Na busca por validação e consistência, indivíduos são motivados a criar obras que podem desafiar o status quo institucional e os valores hegemônicos heteronormativos. Entretanto, nem sempre esse movimento criativo é de subversão, mas sim de afirmação da heteronormatividade. Além disso, a disputa e a busca por reconhecimento muitas vezes são entendidas como empoderamento, mas o termo que era originalmente relacionado à ``dar poder aos sujeitos'' \citet{Grohmann2018}, principalmente a grupos excluídos sistematicamente, hoje também é relacionado à comoditização das experiências das pessoas.
%
%
Esses discursos distintos e até antagônicos são baseados nos horizontes de possibilidades configurados pelas narrativas canônica, aquilo que Jenkins chama ``meta-texto''~\cite{jenkins1992textual}. O meta-texto é uma expectativa dos fãs baseada em seus interesses e desejos particulares sobre a obra. As disputas narrativas são configuradas através de movimentos que buscam a qualificação de determinadas visões a respeito do mundo da história, como o apelo à autoridade do autor, valorização de fragmentos da história canônica em detrimento de outros e discordâncias técnicas. São diversos os caminhos utilizados pelos fãs na busca por legitimação. As interpretações mudam constantemente e são frequentemente discutidas em fóruns e transformadas em trabalhos autorais como \textit{fanfics}, \textit{fanarts} e \textit{fanvideos} (vídeos de fãs), que compõe o universo do ``fanom''\footnote{ ``Fanon'' ou ``fânone'' é a junção da palavra ``canon'' com ``fã'', assinalando a constituição de um universo ficcional em relação paralela à obra oficial, a partir de trabalhos de fãs.}.

\section{O Fandom de Warcraft}

\textit{World of Warcraft} é um jogo extremamente popular, tanto em termos comerciais quanto na academia. Existe um leque imenso de trabalhos a respeito dele, sobre seus aspectos colonialistas e imperialistas, comunidades de jogos, histórias, e mais. Começando pelo prisma do gênero, é difícil aferir com precisão a demografia feminina do jogo, porém os relatos de \citet{Nardi2009} revelam uma grande variedade de jogadores, tanto em relação ao gênero, quanto à escolaridade e profissão; enquanto o trabalho de \citet{Veen2017} mostra uma parcela relevante de 56,9\% de jogadoras ativas. Além disso, um dos maiores websites brasileiros com informações sobre o jogo é o WoWGirl\footnote{WoWGirl. Disponível em \url{https://www.wowgirl.com.br/}.}. Um blog criado em 2008 que se tornou referência no Brasil, sendo composto por uma equipe de maioria feminina~\cite{oliveira2018sexism}.

\citet{sherlock2013happens} trabalhou em cima da retórica de sexualidades \textit{queer} no fandom de \textit{World of Warcraft}. Em seu trabalho, investigou de que maneiras os jogadores e fãs \textit{queer} de WoW estão fazendo um trabalho de ativismo. A pesquisa confronta as expressões no jogo de identidade sexual e performance de gênero com as expressões do fandom \textit{queer}. Ele argumenta que as dinâmicas de jogos \textit{queer} e produção de fãs podem ser instrutivas em conversas sobre retórica e composição, e abrir caminhos para pensarmos sobre identidades e corpos, a expressão de letramentos sexuais e novas paisagens de visibilidade \textit{queer} e ativismo. \citet{sherlock2013happens} aponta como o servidor Proudmoore é conhecido como o ``servidor LGBTQI+'', abrigando uma abundantemente quantidade de guildas amigáveis ao público LGBTQI+. É lá também onde ocorria a anual \textit{Proudmoore Pride Parade}, um evento virtual que simulava paradas gays da realidade, constituindo uma forma de ativismo digital desse público. Mesmo que inicialmente WoW não pressuponha uma cultura LGBTQI+, há uma potência de ação dos jogadores no mundo de jogo, que expressam suas identidades através de seus avatares e recursos virtuais \cite{goulart2012proudmoore}.


Voltando nosso olhar à narrativa de WoW e as disputas do fandom, lançamos mão de alguns estudos anteriores para nos situarmos. Primeiramente, é importante compreender como o próprio cânone do jogo está para além do jogo -- \citet{sbgames2019transmidiawarcraft} analisaram, por meio da cultura participativa e teoria da cibercultura, a narrativa transmídia de alguns jogos (dentre eles WoW) e as ações de fãs, com o objetivo de descrever as estratégias de expansão do universo narrativo e entender como se configuram as narrativas transmidiáticas de jogos digitais. O mapeamento transmídia do universo de Warcraft feito por eles inventariou uma vasta coleção de obras dispersas em diferentes mídias.


Nessa vertente, \citet{sbgames2018aventuraazeroth} estudam a construção de sentidos no universo de World of Warcraft. Eles observaram que os jogos digitais, em particular o WoW, oferecem aos jogadores experiências que não se limitam apenas a uma curiosidade limitada a uma relação tela-jogador, expandindo-se e integrando-se a uma cultura à qual os sujeitos se assimilam construindo e negociando sentidos. Desse modo, os jogadores são extremamente dependentes de outras mídias, comunidades e relações externas ao software para compreenderem os arranjos narrativos de Warcraft e experienciarem o jogo.
\citet{messias2018gambiarra} fez um relato etnográfico realizado em comunidades virtuais do jogo \textit{World of Warcraft} para elucidar como tecnologias digitais como o videogame propiciam reflexões sobre os conceitos de gambiarra e antropofagia. Ele mostra como os mods criados de  WoW estimulam relações de acesso, apropriação tecnológica, capacitação cognitiva, estímulo a criatividade, entre outros. É interessante observar a força política com que a comunidade de fãs se mobiliza para criar e moldar o jogo à sua vontade e preferências através de modificações que alteram a própria jogabilidade, código e percepção dos usuários, e até a criação de servidores paralelos aos oficiais da empresa. 

Mais recentemente, a Blizzard (re)lançou o \textit{World of Warcraft Classic}, uma cópia do jogo original tal qual ele era em seu lançamento em 2004. \citet{schulz2020you} analisa a cultura dos fãs do jogo, explorando as motivações, reações e razões que  dão para jogar WoW Classic como uma forma de mídia nostálgica, estudando assim o fandom através das lentes da ciberetnografia. Os resultados mostram que a popularidade em torno de WoW Classic não se deve apenas ao jogo em si, mas sim à ideia de que a concepção dele foi um passo na direção certa ao dar aos fãs a sensação de agência e voz no desenvolvimento futuro do jogo. Os pesquisadores fornecem um exemplo de como fãs e consumidores podem influenciar as decisões de uma empresa, escolhendo se querem ou não comprar o produto e consumir e qual é esse produto.


Nosso trabalho funciona de forma complementar a esses estudos sobre o fandom de Warcraft. Enquanto grande parte deles focam no jogador e sua relação com o jogo, nós investigamos a produção dos fãs (\textit{fanarts} e \textit{fanfics}) de modo a mapear as disputas narrativas provenientes dos interesses do público e da empresa, considerando as relações de gênero que emergem dessa textualidade.
\section{Metodologia}

Ao trabalhar com material predominantemente digital, principalmente obras populares e com grande público, como WoW e seus jogadores, lidamos com uma abundância de dados disponíveis nos mais variados formatos e redes online. Embora não seja possível cobrir todo esse conteúdo disponível na web, neste trabalho buscamos fornecer um panorama sobre a recepção dos fãs e sua produtividade em relação ao jogo. Para isso, fizemos uma análise de caráter quantitativo em duas das principais plataformas de produções de fãs (AO3 e dA), a fim de mapear as narrativas e representações produzidas e reiteradas pelo fandom. A seguir, entraremos em detalhes em como cada um desses corpus foi coletado e construído.

\subsection{Coleta de Dados de \textit{Fanfics}}

\textit{Fanfics} ou \textit{fanfictions} são uma categoria de produção textual muito popular nos fandoms e na internet. São vários os locais onde um usuário pode publicar seu trabalho, desde blogs pessoais, até grandes repositórios. Para as coletarmos foi escolhido o site Archive of Our Own, que oferece um ``local central de hospedagem, não comercial e sem fins lucrativos, para trabalhos de fãs, usando um software de código aberto de arquivamento''\footnote{Disponível em: \url{https://archiveofourown.org/about}.}.
O AO3 é um dos maiores sites de \textit{fanfic} da internet, hospedando mais de 6 milhões de trabalhos e premiada várias vezes\footnote{Disponível em: \url{https://www.b9.com.br/112745/site-de-fanfics-archive-of-our-own-ganha-premio-hugo-de-literatura/}.}.

Uma \textit{fanfic} é exibida no site do AO3 contendo várias informações associadas disponíveis para cada trabalho. Junto ao título da \textit{fanfic} e apelido do autor, além do fandom ao qual o trabalho está vinculado, existe um conjunto gráfico de 4 símbolos que sumariza algumas informações gerais da obra: \textit{Rating} (faixa etária recomendada), \textit{Relationship} (tipo de relacionamentos), \textit{Warnings} (advertência de conteúdo sensível) e \textit{Status} (estado de desenvolvimento do trabalho). O site também fornece \textit{Warnings} (alertas), lista de \textit{Characters} (personagens) presentes na história e uma lista de \textit{Additional Tags} (etiquetas adicionais) atribuídas à obra. Vale ressaltar que esses dados são fornecidos pelo próprio autor, que pode optar por incluí-los ou não na obra.
%
%
Todas essas categorias e informações presentes nos trabalhos nos ajudam a compreender melhor o universo do \textit{fanfics}. Ao observar como usuários utilizam esse sistema e algumas tendências que emergem nesse meio, temos um vislumbre de aspectos gerais do fandom que serão úteis para análises mais aprofundadas a respeito das obras. Para analisar esse cenário, foi utilizada uma metodologia de coleta de dados consistindo em duas etapas:

\begin{enumerate}
    
    \item Listagem e coleta de todas as \textit{fanfics} através das especificações fornecidas para busca de \textit{fanfics} de World of Warcraft;
    
    \item Coleta individual dos dados de cada uma das \textit{fanfics}, com o texto e metadados fornecidos pelos autores. 

\end{enumerate}

Como a plataforma do AO3 armazena as \textit{fanfics} de forma estruturada, a interface conta com um complexo sistema de busca, que permite a filtragem por tags de referência. É possível buscar trabalhos por tags, data, idioma, incluir ou excluir \textit{crossovers}, buscar por termos, autor, título, etc.

Neste trabalho, realizamos uma busca dentro do AO3 de todos os trabalhos catalogados pela plataforma como pertencentes ao fandom ``\textit{World of Warcraft}''. 
Todos os resultados obtidos nesta filtragem são coletados com os metadados de cada um. Para esta coleta, foi utilizada uma biblioteca Python chamada \textit{ao313}\footnote{Disponível em: \url{https://github.com/alexwlchan/ao3}.}, uma API não-oficial que coleta dados de \textit{fanfics} do site AO3 através de um processo de raspagem de dados. 
Com esse método foi possível coletar todos os 3999 textos listados no dia 28 de junho de 2019 pelo site e extraímos as seguintes informações: título, autor, data de publicação, classificação indicativa, \textit{tags} diversas, personagens, relacionamentos, \textit{kudos}, comentários, e até o próprio texto da \textit{fanfic}.

\subsection{Coleta de Dados de \textit{Fanarts}}

Semelhante às \textit{fanfics}, \textit{fanarts} são postadas pelos artistas em diversas plataformas online. Para buscarmos e coletarmos nosso material optamos pelo site DeviantArt (dA). Ele é especificamente voltado para o público artístico e possui importância histórica sobre os modos como artistas se organizam na internet~\cite{ables_2019}, constituindo uma das maiores comunidades artísticas online, com mais de 48 milhões de usuários registrados e atraindo mais de 45 milhões de visitantes por ano.

O dA hospeda uma quantidade diversa de tipos de arte, desde fotografias até arte digital, oferecendo ferramentas de organização e exibição para os artistas. Ao publicar uma arte, o usuário pode adicionar um texto descritivo, até 30 \textit{tags} e marcar o conteúdo como adulto (``\textit{Mature Content}''). Após a submissão, cada obra passa a contar também com estatísticas sobre quantos comentários recebeu, quantas vezes foi marcada como favorita e número de visualizações. 



Além disso, cada autor no site possui uma página de perfil que consta com: uma galeria para seus trabalhos, seção com informações sobre o artista, seção de favoritos, uma área para postagens, e até mesmo uma loja. Para este trabalho, focamos nas informações pessoais do usuário presentes ali como: nome, quantidade de \textit{watchers} (seguidores) que possui, quantas visitas o perfil já recebeu, o total de artes publicadas, total de comentários e total de favoritos. A profissão e área que a pessoa atua também constam nos perfis, além da data de aniversário, país e idade da conta de usuário. O artista também pode visibilizar o seu gênero através do uso dos pronomes exibidos no perfil: ``\textit{\textbf{They/Them}}'', ``\textit{\textbf{She/Her}}'' e ``\textit{\textbf{He/Him}}''. A informação pode estar oculta pelo termo ``\textit{unknown}''.


Para coletar estas informações e explorar aspectos como quem produz no fandom, quais são essas produções e como ocorre a disputa narrativa sobre WoW, decidimos buscar por \textit{fanarts} de Jaina Proudmoore, uma das personagens femininas mais populares do jogo, presente desde suas primeiras versões e que vem ganhando maior protagonismo nas últimas expansões. Deste modo, nos atemos a uma narrativa específica (a de Jaina) e podemos observar como os esforços por melhores representações femininas ocorrem no jogo e no fandom. 

Para automatizar e coletar os dados, primeiro foi necessário filtrar as artes relacionadas ao tema. O site do dA permite pesquisas com termos, então utilizamos o sistema para procurar por todas obras com a expressão ``Jaina Proudmoore''. Vale mencionar que, para que todas obras sejam listadas nos , é necessário fazer \textit{login} com uma conta e habilitar a exibição de conteúdo adulto, dado que essas obras são excluídas da busca por padrão. Com a busca foram retornadas 2281 artes relacionadas à personagem.



Até onde sabemos, não existem ferramentas oficiais (códigos, scripts ou mesmo uma API (Application Programming Interface) para filtrar e coletar dados do dA. Portanto, para efetuar a coleta, foi implementada e realizada uma raspagem de dados baseada na estrutura das páginas HTML tanto das \textit{fanarts} como de seus usuários autores. Mais detalhadamente, este processo é realizado em três etapas distintas:

\begin{enumerate}
    
    \item Coleta de todos os links listados pelo resultado da busca; 
    
    \item Coleta do conteúdo de cada link, com todas as informações da obra;
    
    \item Coleta da página de perfil do artista para cada uma das artes, com as informações disponíveis de cada um. 

\end{enumerate}


Utilizamos scripts desenvolvidos em linguagem Python com uso da biblioteca \textit{Selenium}\footnote{Disponível em \url{https://selenium-python.readthedocs.io/}}, que permite manipulação e gerenciamento de dados de páginas através de seu endereço URL. Ao final, foi possível obter um conjunto de dados com as 2281 obras relacionadas ao fandom de WoW e à Jaina Proudmoore, composto por diferentes estilos de arte, como pintura digital, 3D, desenho tradicional, fotografias, \textit{cosplays}, capturas de tela de jogo, entre outras.

\section{Mergulhando no Conteúdo dos Fãs}

\begin{table}[t]
    \centering \hspace{-0.2cm}
    \scriptsize
    \caption{Resumo dos dados coletados de fãs sobre \textit{fanfics} e \textit{fanart}.}
    \label{tab:dados}
    \begin{tabular}{|c|c|c|c|c|c|} 
    \hline 
    \textbf{Tipo} & \textbf{Plataforma}                                           & \textbf{\begin{tabular}[c]{@{}c@{}}Total de\\ Trabalhos\end{tabular}} & \textbf{\begin{tabular}[c]{@{}l@{}}Total de\\ Usuários\end{tabular}} & \textbf{\begin{tabular}[c]{@{}c@{}}Data de\\ Coleta\end{tabular}} & \textbf{\begin{tabular}[c]{@{}c@{}}Termo de \\ Busca\end{tabular}} \\ \hline
    \textit{Fanarts}       & DeviantArt                                                    & 2281                                                                   & 1385                                                                  & 27/03/2020                                                        & \begin{tabular}[c]{@{}c@{}}Jaina\\Proudmoore\end{tabular}                                                   \\ \hline
    \textit{Fanfics}       & \begin{tabular}[c]{@{}c@{}}Archive Of\\  Our Own\end{tabular} & 3999                                                                   & Indisponível                                                                  & 28/06/2019                                                        & \begin{tabular}[c]{@{}c@{}}World of\\Warcraft\end{tabular}                                                  \\ \hline
    \end{tabular}
    \vspace{-0.3cm}
\end{table}

\begin{figure}[t]
    \centering 
    \includegraphics[width=0.85\linewidth]{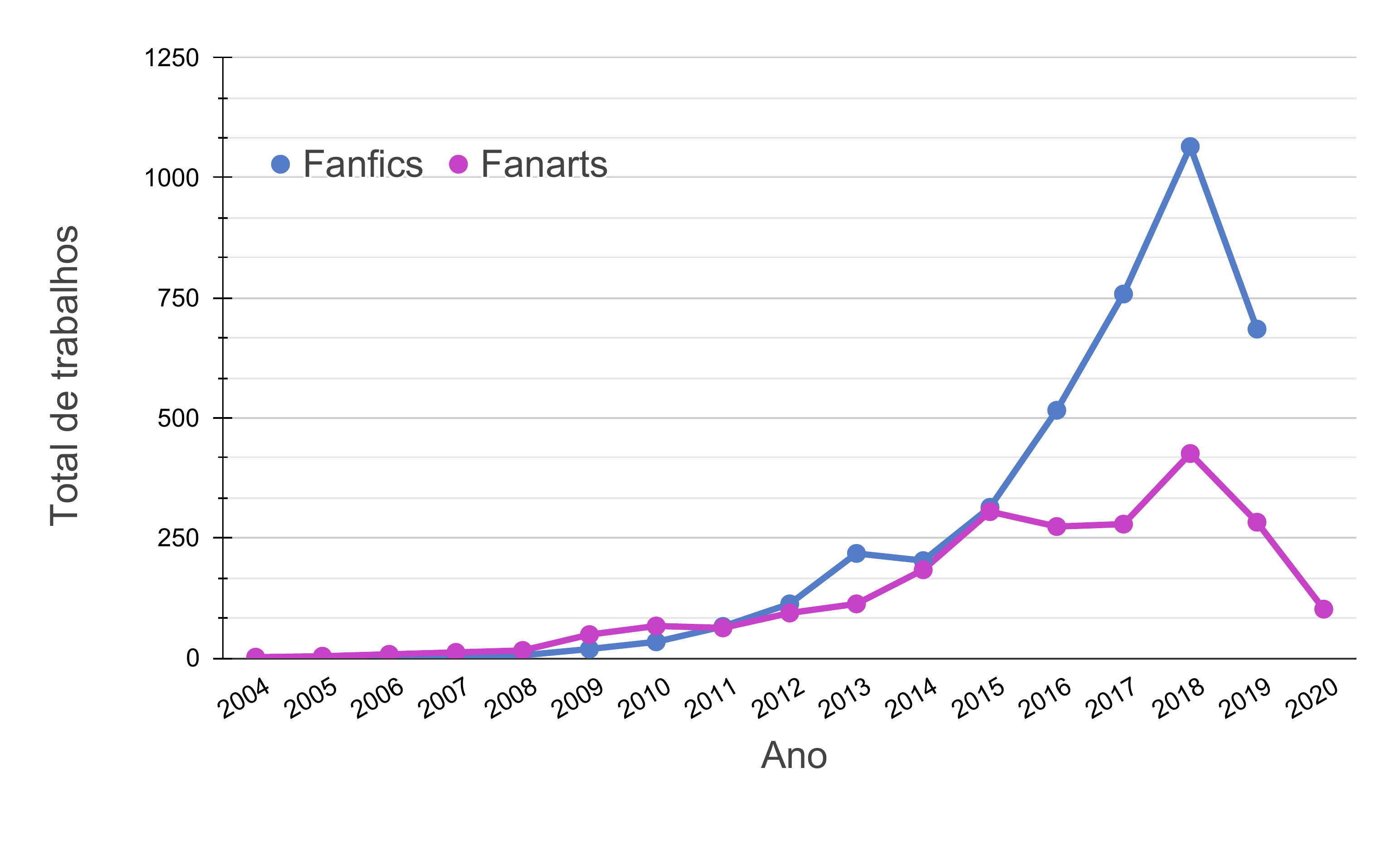}
    \vspace{-0.7cm}
    \caption{Número de trabalhos de \textit{fanarts} e \textit{fanfics} coletados por ano no fandom de\textit{ World of Warcraft}.}
    \vspace{-0.4cm}
    \label{fig:year}
    
\end{figure}


A Tabela \ref{tab:dados} resume a amostragem. Como o dA e o AO3 são plataformas disponíveis há vários anos e WoW surgiu em 2004, podemos traçar um histórico de produção relacionado à WoW e Jaina Proudmoore ao longo de todo o período de vigência do jogo, como mostrado na Fig.~\ref{fig:year}. Embora vários outros fatores influenciem na quantidade de trabalhos disponíveis, como a democratização do acesso a essas plataformas e a popularização do usuário como produtor de conteúdo em vez de mero consumidor online, é notável a popularidade do jogo desde seu lançamento, com um pico em ambas plataformas em 2018, ano de lançamento da expansão \textit{Battle for Azeroth}.

A seguir, analisamos as coleções explorando as informações disponíveis sobre cada categoria de mídia e plataformas escolhidas, começando por um olhar do universo geral do fandom de \textit{World of Warcraft} no AO3 e depois nos aprofundando nas \textit{fanarts} sobre Jaina Proudmoore no dA.

\subsection{Fanfics}\label{sec:fanfic_data}

\begin{figure}[t!]
    \centering
    \begin{minipage}[h!]{0.49\linewidth}
        \includegraphics[width=1.0\linewidth]{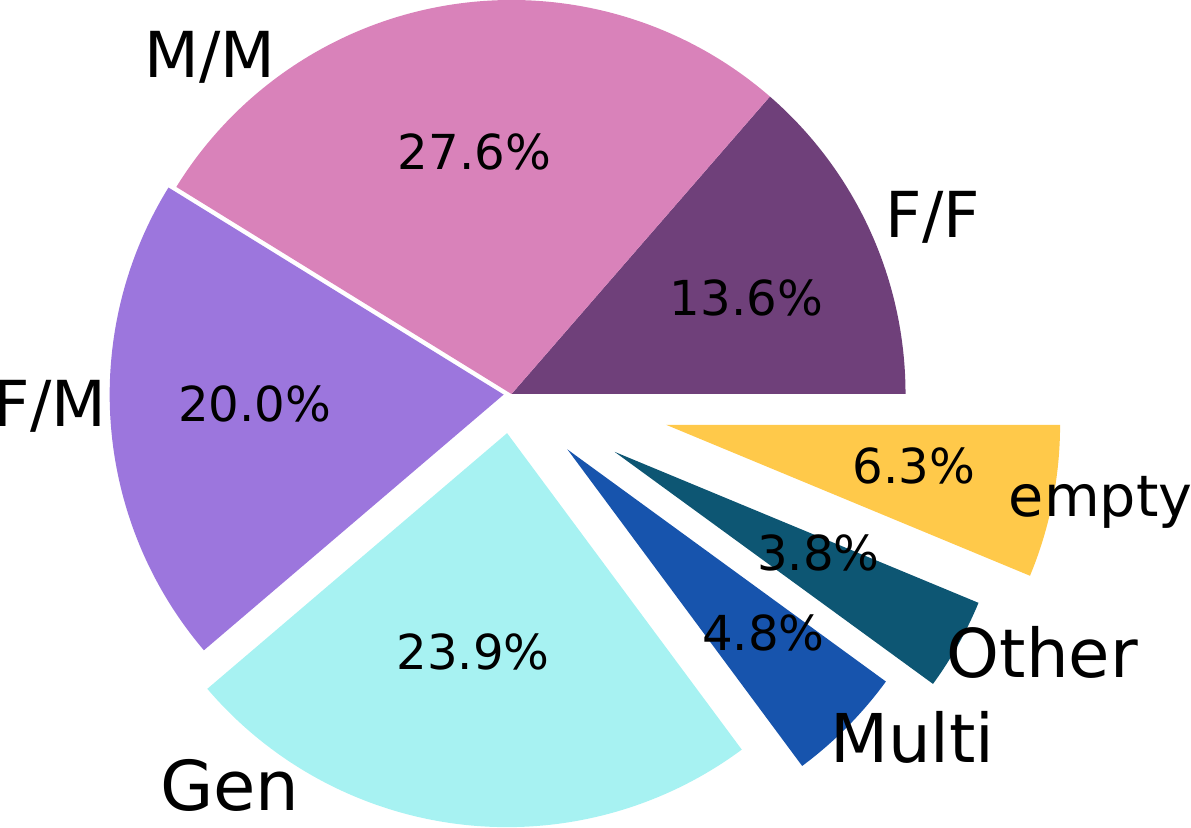}
        \caption{Gênero de relacionamento das \textit{fanfics} de Warcraft.}  \label{fig:relationship}
    \end{minipage}
    \hfill
    \begin{minipage}[h!]{0.47\linewidth}
        \includegraphics[width=1.0\linewidth]{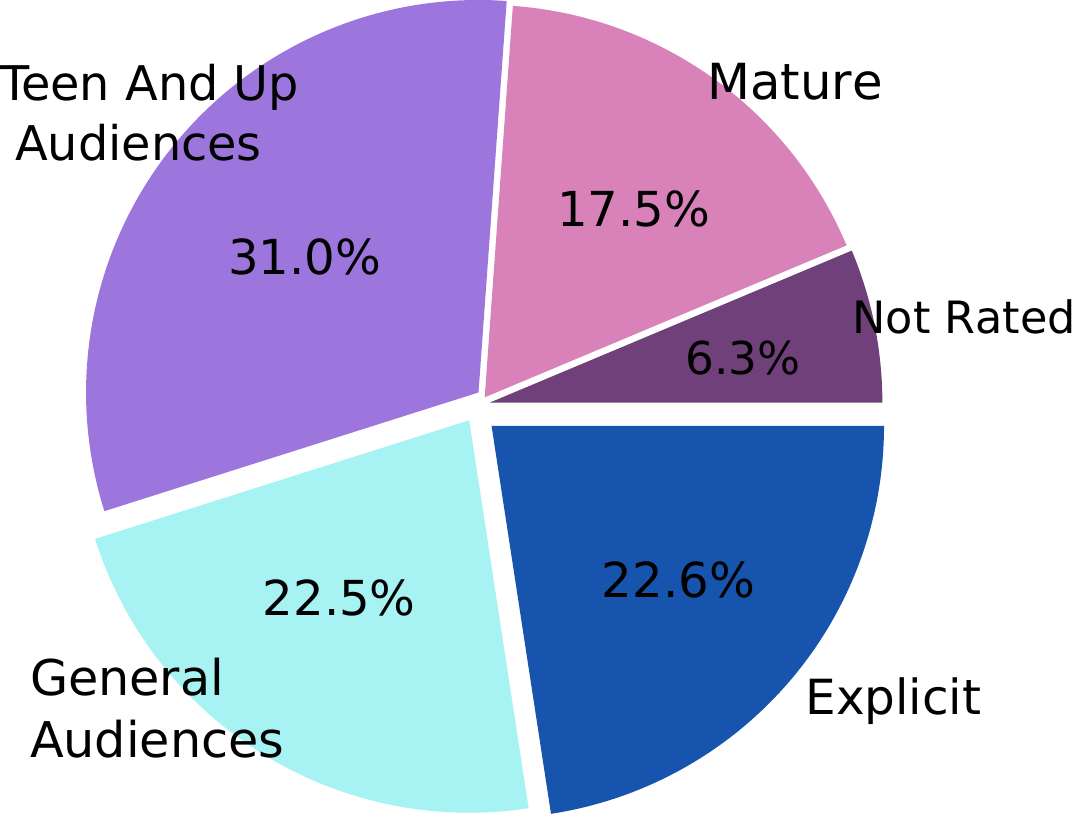}
        \caption{Classificação de conteúdo das \textit{fanfics} de Warcraft no AO3.}
        \label{fig:rating}
    \end{minipage}
    \vspace{-0.3cm}
\end{figure}

\begin{figure}[]
    \centering 
    \includegraphics[width=0.72\linewidth]{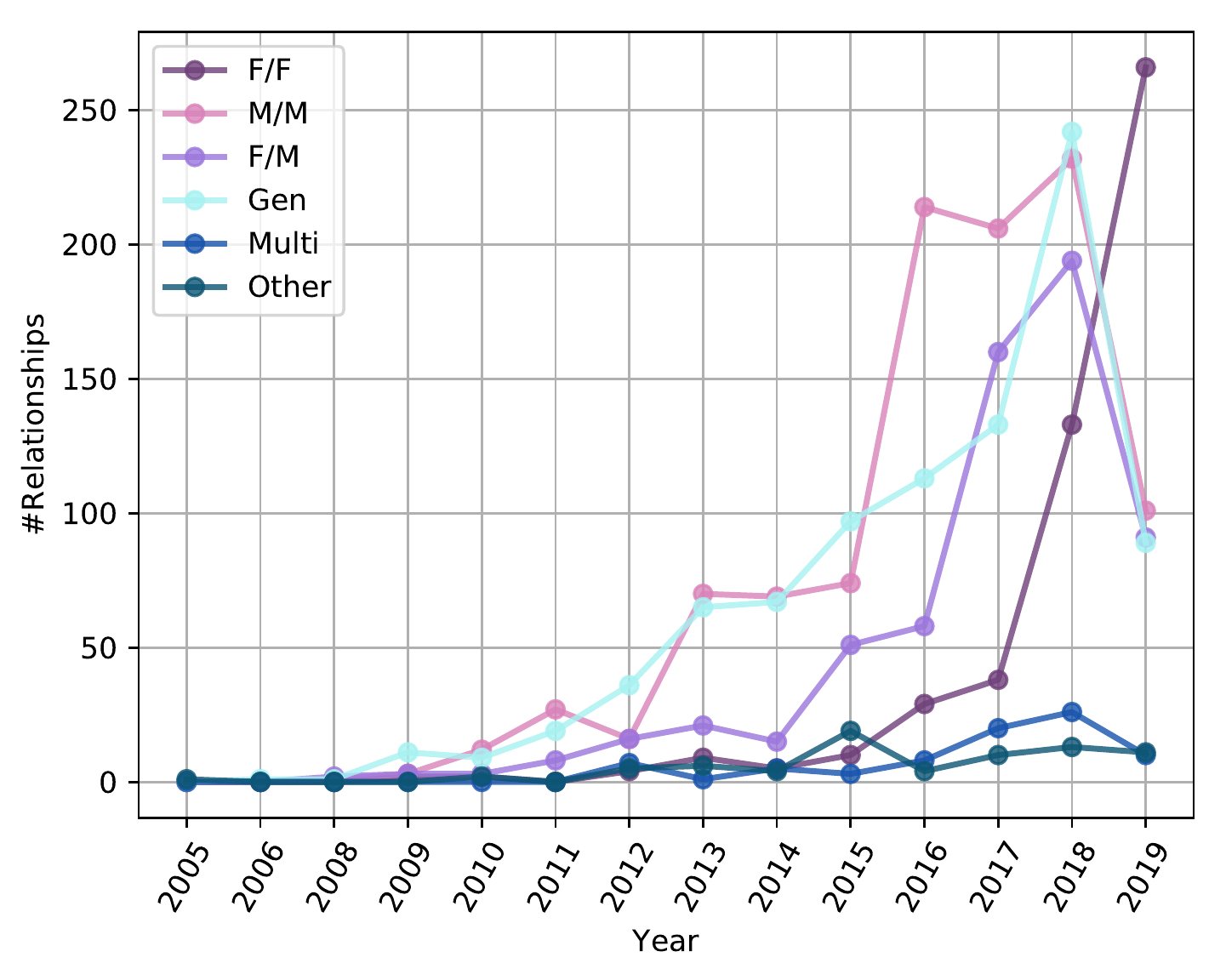}
    \vspace{-0.35cm}
    \caption{Total de gêneros dos relacionamentos por ano nas \textit{fanfics} de World of Warcraft coletadas.}
    \label{fig:relationship_year}
    \vspace{-0.4cm}
\end{figure}

As primeiras informações que observamos são as de ``\textit{relationship}'' e ``\textit{rating}'' das \textit{fanfics}. A Fig.~\ref{fig:relationship} mostra a composição geral de todos os textos coletados quanto a categoria dos relacionamentos entre personagens. Percebemos que 23.9\% das histórias não possuem relacionamentos românticos ou sexuais. Nas que possuem, os relacionamentos ``M/M''~(Homem/Homem) são predominantes (27.6\%), seguidos por 20\% de ``F/M''~(Mulher/Homem) e 13.6\% de ``F/F''~(Mulher/Mulher). Podemos supor através disso que há um protagonismo maior de personagens masculinos nas histórias, expresso pela maioria de relacionamentos homossexuais entre homens. Junto à presença de relacionamentos lésbicos, isso também evidencia a presença de narrativas não-heterossexuais no fandom, colocando em pauta que o público do jogo também é composto por pessoas LGBTQI+ ou indivíduos com interesses nessas histórias. 
Por outro lado, ao organizarmos essa distribuição por ano (Fig.~\ref{fig:relationship_year}), observamos que a partir de 2018 a maioria de relacionamentos postados na plataforma é ``F/F'' (Mulher/Mulher). Essa popularidade coincide ao aumento da participação das mulheres na história do jogo nos últimos anos através de personagens como Jaina Proudmoore e Sylvana Correventos; ambas protagonizam curtas da série \textit{Warbringers} e passaram a ocupar papéis de liderança em destaque no jogo após 2018.

Uma grande porcentagem dos trabalhos possui o rótulo Maduro e Explícito (17,5\% e 22,6\% respectivamente), como visto na Fig.~\ref{fig:rating} sobre a classificação etária dos trabalhos, sinalizando que quase metade das obras possuem temas adultos, como violência e sexo. Isso, juntamente à prevalência de casais ``M/M'' nas \textit{fanfics} e a ascensão de ``F/F'', acompanha uma tendência geral da plataforma: metade dos trabalhos do AO3 tem casais masculinos e um terço está assinalado como maduro ou explícito~\cite{Massey2019}. Um censo realizado no AO3~\cite{lulucensoao3}
em 2013 apurou que a maioria dos autores na plataforma são mulheres (90,3\%) e pessoas \textit{queer} (7.3\%). Historicamente, os estudos de fãs abordam as \textit{slash fanfictions}, histórias de cunho sexual sobre relacionamentos homossexuais, normalmente produzidas por mulheres. Essas obras são criadas sob lógica de \textit{gift economy}, possuem um caráter afetivo e são associadas ao desejo de explorar a sexualidade e identidades em espaços seguros~\cite{Massey2019}. Desse modo, enxergamos no fandom de WoW uma tendência de produção de \textit{slash fanfictions} com ênfase na exploração da sexualidade.

\begin{figure}[!t]
    \centering
    \begin{minipage}[h!]{.8\linewidth}
        \includegraphics[width=1.0\linewidth]{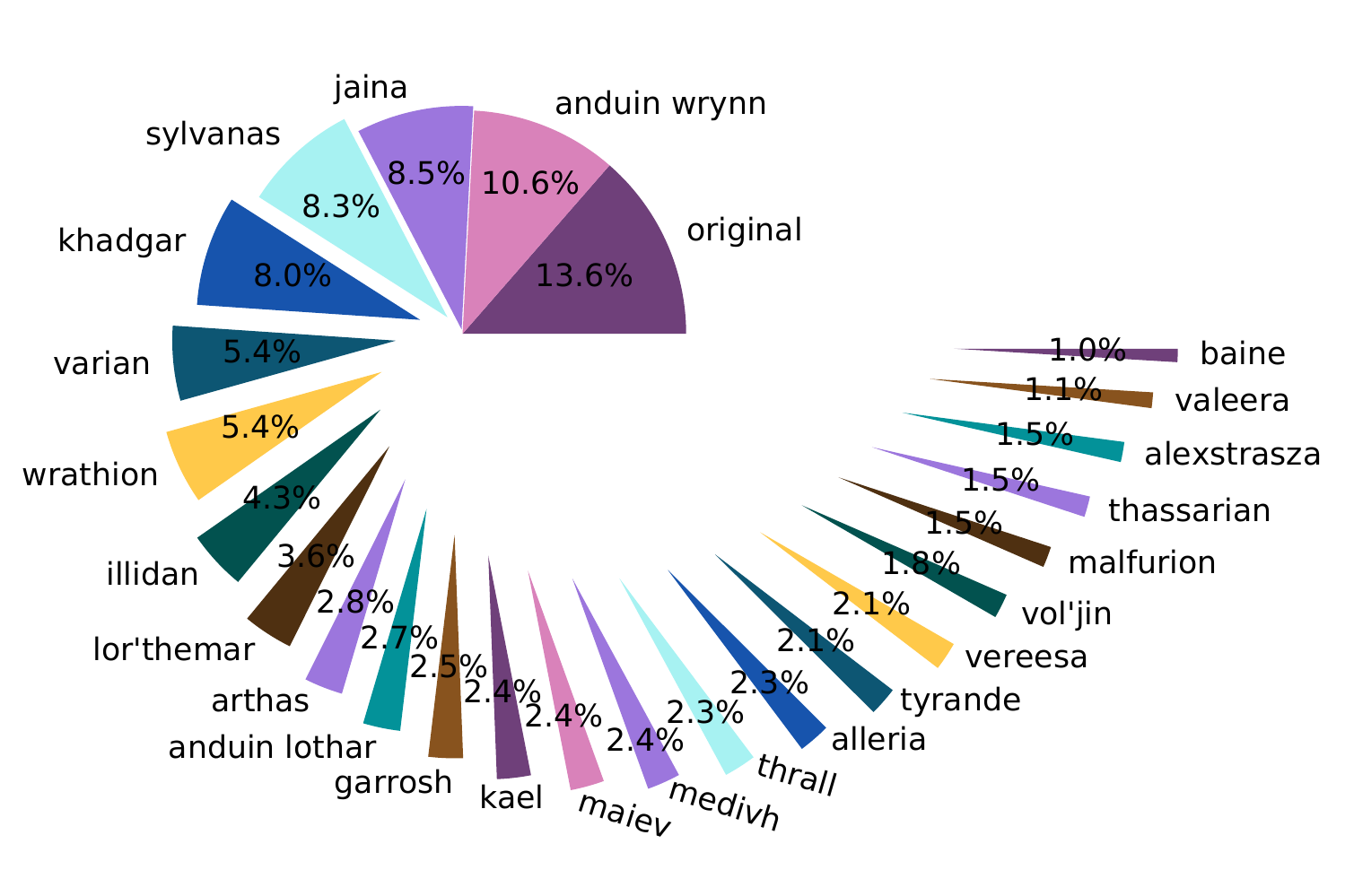}
        \vspace{-0.90cm}
        \caption{Presença dos personagens dentre as \textit{fanfics} de\\ Warcraft.}
        \label{fig:characters}
    \end{minipage}
    \begin{minipage}[h!]{.999\linewidth}
        \includegraphics[width=1.0\linewidth]{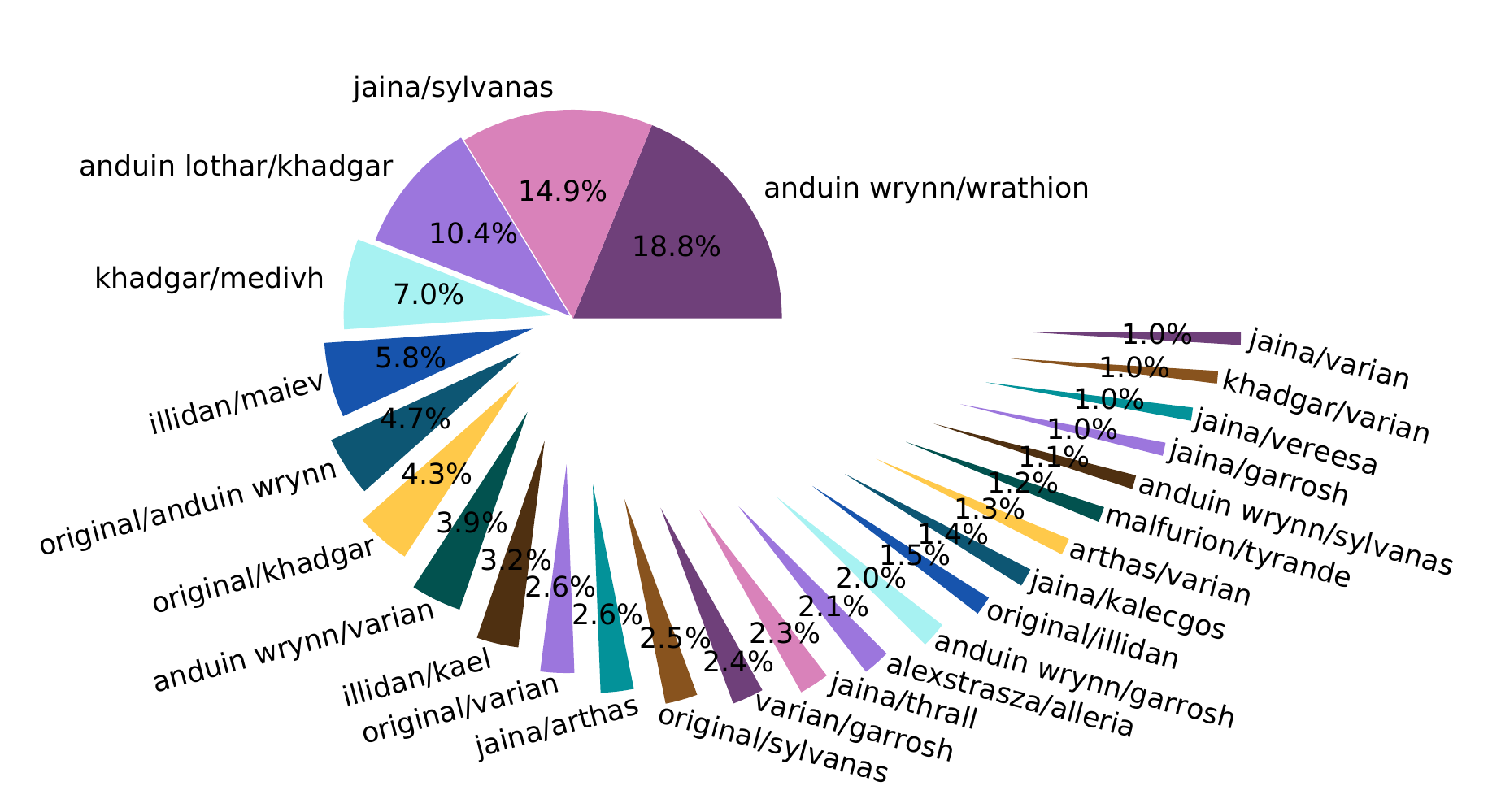}
        \vspace{-0.75cm}
        \caption{Presença de relacionamentos dentre as \textit{fanfics} de Warcraft.}
        \label{fig:partners}
    \end{minipage}
    \vspace{-0.65cm}
\end{figure}

Nos gráficos das Fig.~\ref{fig:characters} e Fig.~\ref{fig:partners}, organizamos a participação das personagens, além de exibirmos os relacionamentos tratados pelo fandom. Jaina e Anduin Wrynn são os personagens mais populares, superados apenas por personagens originais. Como WoW é um MMORPG, é comum que as pessoas utilizem seus personagens de jogo em \textit{fanfics} -- isso é interessante visto que muitas vezes o personagem também é um avatar do jogador.


O relacionamento mais frequente é romântico e homossexual, entre Anduin Wrynn e Wrathion; logo em seguida temos o relacionamento lésbico de Jaina e Sylvana. Enquanto Anduin e Wrathion são amigos na história canônica, Jaina e Sylvana são rivais e não se gostam. Essa inimizade possui um potencial dramático que pode ser explorado em histórias; inclusive há \textit{tags} comumente relacionadas a isso, como ``\textit{enemies to lovers}'' e ``\textit{slow burn}''. Além disso, ``jaina/sylvanas'' é o único casal ``F/F'' da lista com personagens canônicos -- o que evidencia a baixa quantidade de personagens femininas no cânone e sugere que o horizonte de expectativas para essa categoria de relacionamento é reduzido.

\begin{figure}[t]
    \centering 
    \makebox[\linewidth][c]{
        \includegraphics[trim={3.1cm 2.75cm 3.1cm 6.50cm},clip, width=1.0\linewidth]{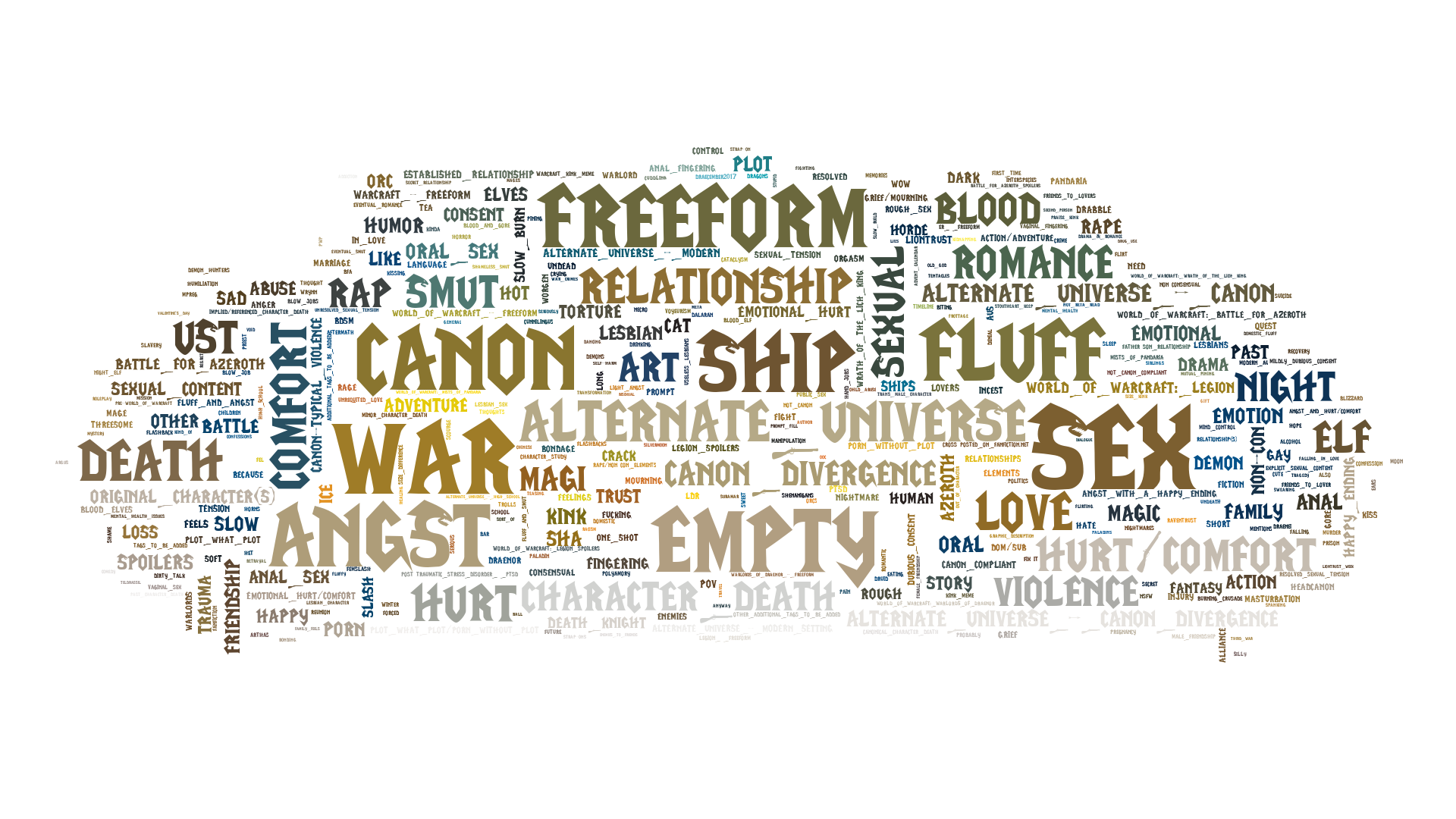}
    }
    \vspace{-1.25cm}
    \caption{ Nuvem de termos com a tags utilizadas nas \textit{fanfics} do fandom de Word of Warcraft coletadas no AO3.}
    \label{fig:wordcloud_tags}
    \vspace{-0.5cm}
\end{figure}

\begin{figure}[]
    \centering 
    \makebox[\linewidth][c]{
        \includegraphics[trim={0 0 01cm 0},clip, width=1.0\linewidth]{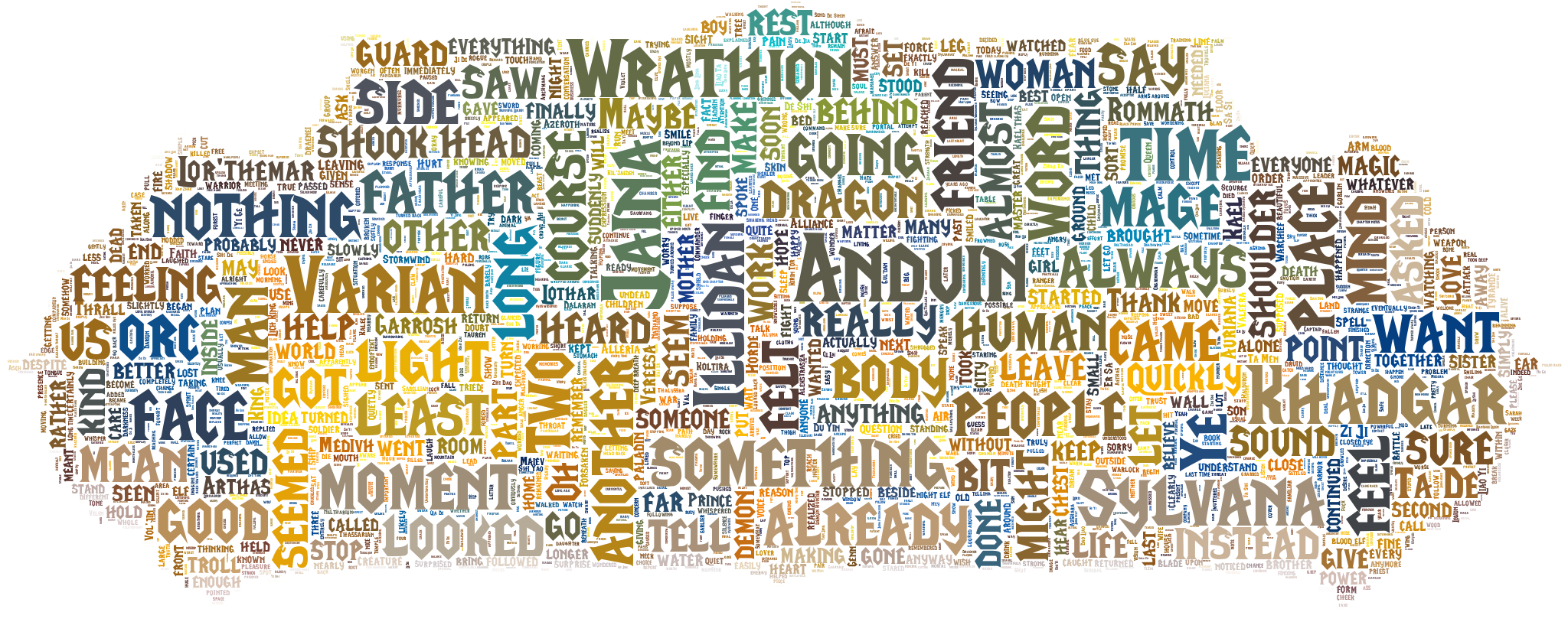}		   	 
    }
    \vspace{-0.75cm}
    \caption{ Nuvem de termos com o corpo de textos de todas \textit{fanfics} do fandom de World of Warcraft coletadas no AO3.}
    \label{fig:wordcloud_fanfic}
    \vspace{-0.5cm}
\end{figure}

Outras formas de encararmos o conteúdo é através das \textit{tags} de categorização e do próprio conteúdo textual das histórias. Uma forma de ter uma ideia geral sobre o conteúdo e popularidade de personagens é através de nuvens de termos. Na nuvem da Fig.~\ref{fig:wordcloud_tags}, podemos ter uma ideia geral do conteúdo e popularidade de personagens no interior dos textos. O nome Anduin possui maior recorrência, entretanto é importante salientarmos que existem dois Anduin em WoW (Anduin Wrynn e Anduin Lothar). Os nomes dos personagens populares do fandom também são comuns nas histórias (Jaina, Khadgar, Wrathion, Sylvana, Varian).
O tema mais frequente entre os trabalhos é ``\textit{angst}'' (angústia),  seguido por ``\textit{fluff}'' (histórias com conteúdos felizes). É curioso que as duas categorias sejam tão populares, pois isso aponta para um fandom com uma produção polarizada ou cujos textos tratam de dicotomias como tristeza/felicidade. 
\textit{Tags} sexuais (\textit{smut}, \textit{oral sex} e \textit{anal sex}) e de romance são bastante populares, deixando visível que os usuários discorrem sobre sexualidade nas histórias. Junto às categorias de ``F/F'' e ``M/M'' isso demonstra que a maioria dos relacionamentos são sexuais, não apenas românticos ou platônicos. Violência também é um tema recorrente e podemos relacionar esse fenômeno à história canônica de WoW, que trata de guerras e conflitos entre povos. Embora seja difícil destacar algum tópico dentre as palavras na Fig.\ref{fig:wordcloud_fanfic}, percebemos a presença de palavras como ``\textit{Dragon}'', ``\textit{Demon}'', ``\textit{Orc}'', ``\textit{Elf}'', ``\textit{Magic}'', que salientam o gênero de fantasia ao qual Warcraft pertence, mostrando que muitas \textit{fanfics} também abordam essa temática. 




\subsection{Fanarts}
\vspace{-0.1cm}
Agora nos atentaremos ao conjunto de dados de fanarts coletadas da plataforma do deviantArt. O dA é uma plataforma para os artistas publicarem seus trabalhos funcionando como uma galeria, dessa forma, abordamos não só os dados das imagens coletadas, mas também as informações disponíveis sobre os artistas que as publicaram.


\begin{figure}[t]
    \centering
    \begin{minipage}[h!]{.48\linewidth}
        \includegraphics[width=0.91\linewidth]{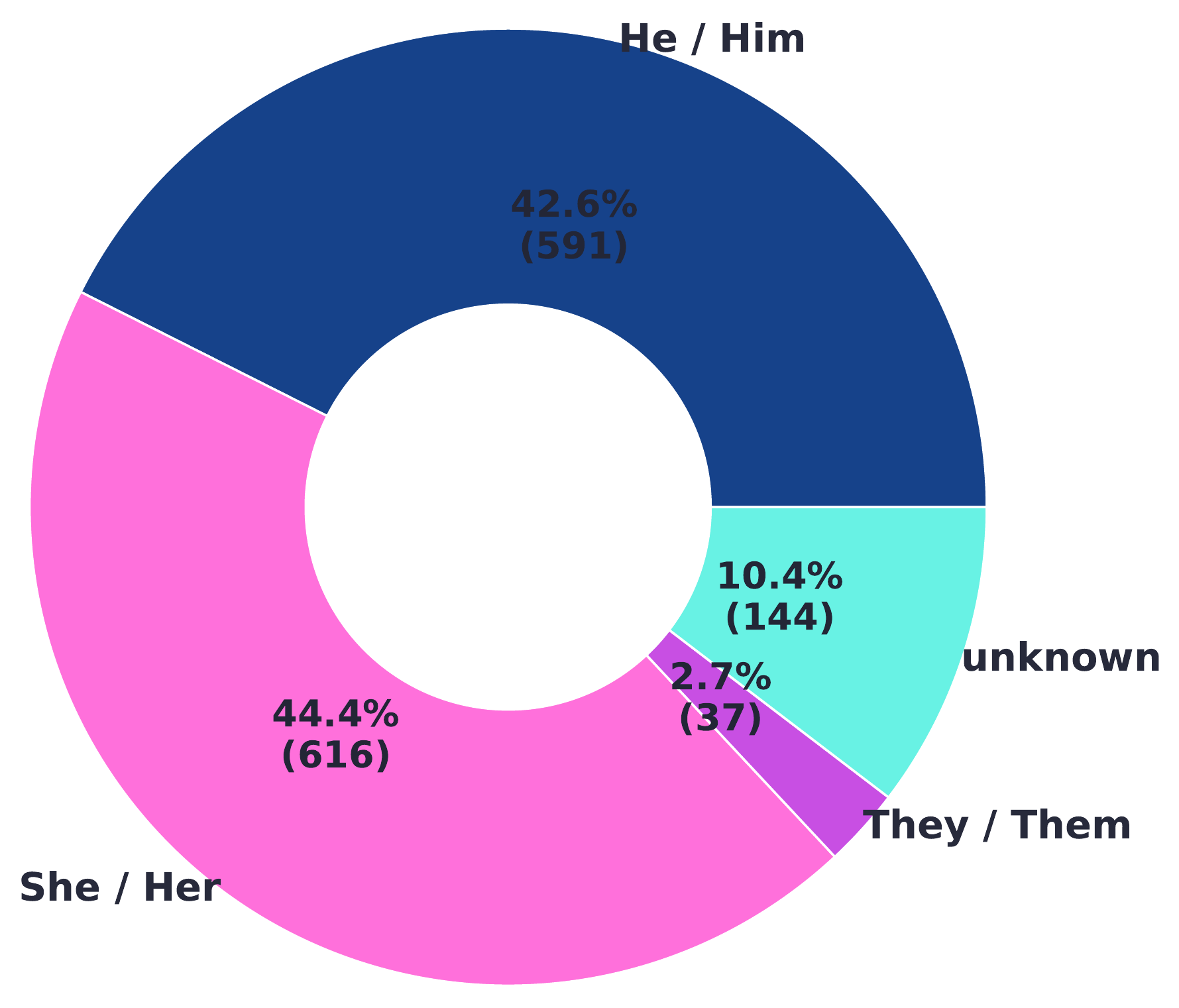}
        \caption{Gênero dos artistas.}\label{fig:artist_gender}
    \end{minipage}
    \hfill
    \begin{minipage}[h!]{.49\linewidth}
        \includegraphics[width=1.00\linewidth]{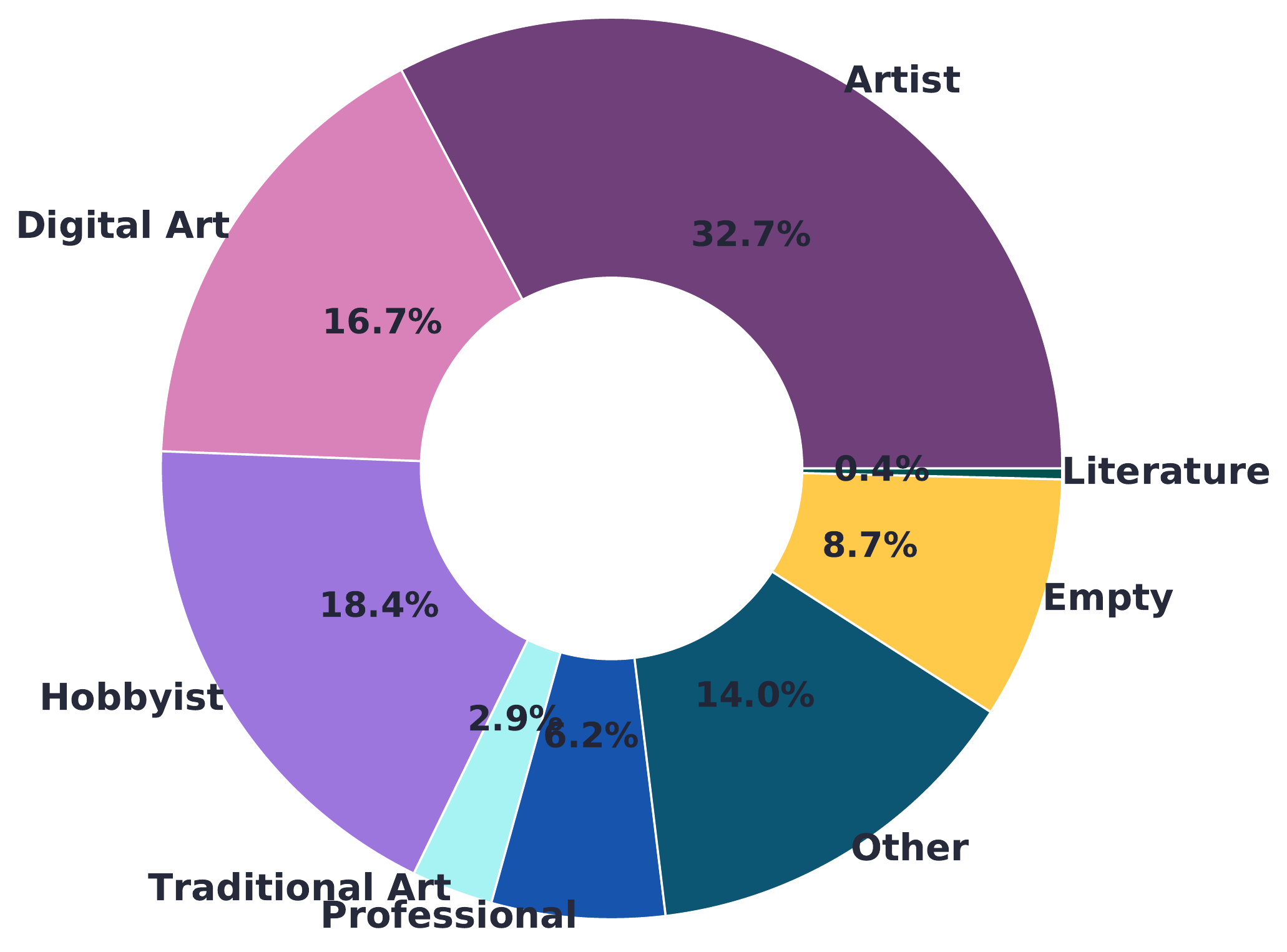}
        \caption{Ocupação e especialidade dos artistas.}\label{fig:artist_occ}
    \end{minipage}
    \vspace{-0.3cm}
\end{figure}

\begin{figure}[!t]
    \centering 
    \includegraphics[width=0.72\linewidth]{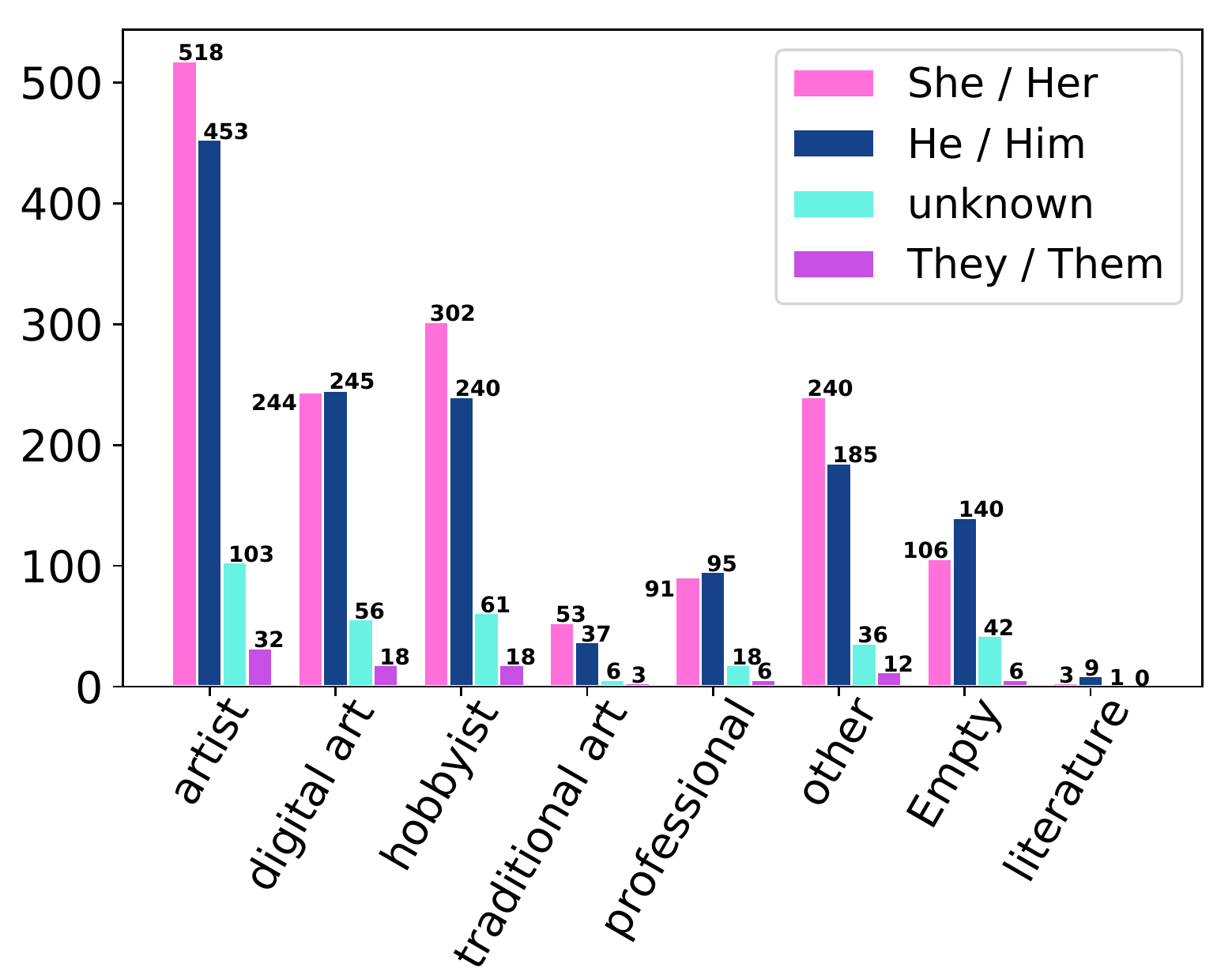}
    \vspace{-0.2cm} 
    \caption{Divisão por gênero das ocupações e especialidades dos artistas das \textit{fanarts} no DeviantArt.}		
    \label{fig:gender_occ}
    \vspace{-0.5cm}
\end{figure}

Começamos observando o gênero dos artistas com a Fig.~\ref{fig:artist_gender}, visto que essa informação é importante, uma vez que nos ajuda a perceber quem publica \textit{fanarts} de WoW na plataforma. Embora seja possível esconder essa informação do perfil, observamos que apenas 10.4\% dos usuários escolheram por não informá-la enquanto a maioria opta por deixar visível. Outra constatação é a distribuição muito parecida entre homens (``\textit{He/Him}'') e mulheres (``\textit{She/Her}''), com uma ligeira maioria de mulheres artistas (42,6\% e 44,4\% respectivamente). Além disso, temos 2.7\% dos usuários, o que significa um total de 38 artistas, se identificando com o pronome ``textit{They/Them}''. É interessante observar que os pronomes neutros simbolizam a participação de pessoas \textit{queer} no fandom, apesar da baixa presença. Esta distribuição difere da que observamos no AO3, onde supomos uma maioria ampla de mulheres escrevendo e mais pessoas \textit{queer} do que homens. Isso nos sugere que a produção de \textit{fanarts} ocorre de maneira diferente da de \textit{fanfics}, podendo haver uma discrepância produzida pelo acesso às ferramentas de criação e plataforma.

Além disso, no perfil do dA, os usuários podem descrever suas ocupações como artistas. Eles podem escolher se seu perfil é profissional, de hobby, ou mesmo estudantil. Também podem especificar sua especialidade dentre artesanato, design, arte digital, arte tradicional, 3D, animação, ou mesmo literatura. A Fig.~\ref{fig:artist_occ} mostra a distribuição da ocupação dos autores das \textit{fanart}. Essa informação pode estar escondida ou indisponível, representado aqui por “\textit{Empty}”. Uma grande quantidade de usuários se apresenta como artista, com mais de 1000 pessoas marcando a opção, o que não é surpreendente, dado que o dA é uma plataforma voltada para esse público. Muitos autores se posicionam como ``\textit{Hobbyist}'' (621 usuários), o que demonstra a predominância da produção de \textit{fanarts} de Jaina como um ato voluntário e prazeroso, o que condiz com a lógica da economia de presente. Em comparação aos usuários que produzem por hobby, existe uma quantidade relevante de pessoas que se identificam como ``\textit{Professional}'' (210 usuários), o que também aponta para um deslocamento das \textit{fanarts} para a esfera do comércio e trabalho.

Na Fig.~\ref{fig:gender_occ}, temos a distribuição de ocupação por gênero, e fica evidente uma diferença quanto a profissionalização. Há uma maioria feminina entre quem cria por hobby, enquanto os profissionais são ligeiramente uma maioria masculina. Esses resultados sugerem que o público feminino do fandom de WoW produz \textit{fanarts} de maneira voluntária, enquanto o masculino está mais ligado ao trabalho e a produção comercial deste tipo de conteúdo.

\begin{figure}[t]
        \center
        \includegraphics[width=0.88\linewidth]{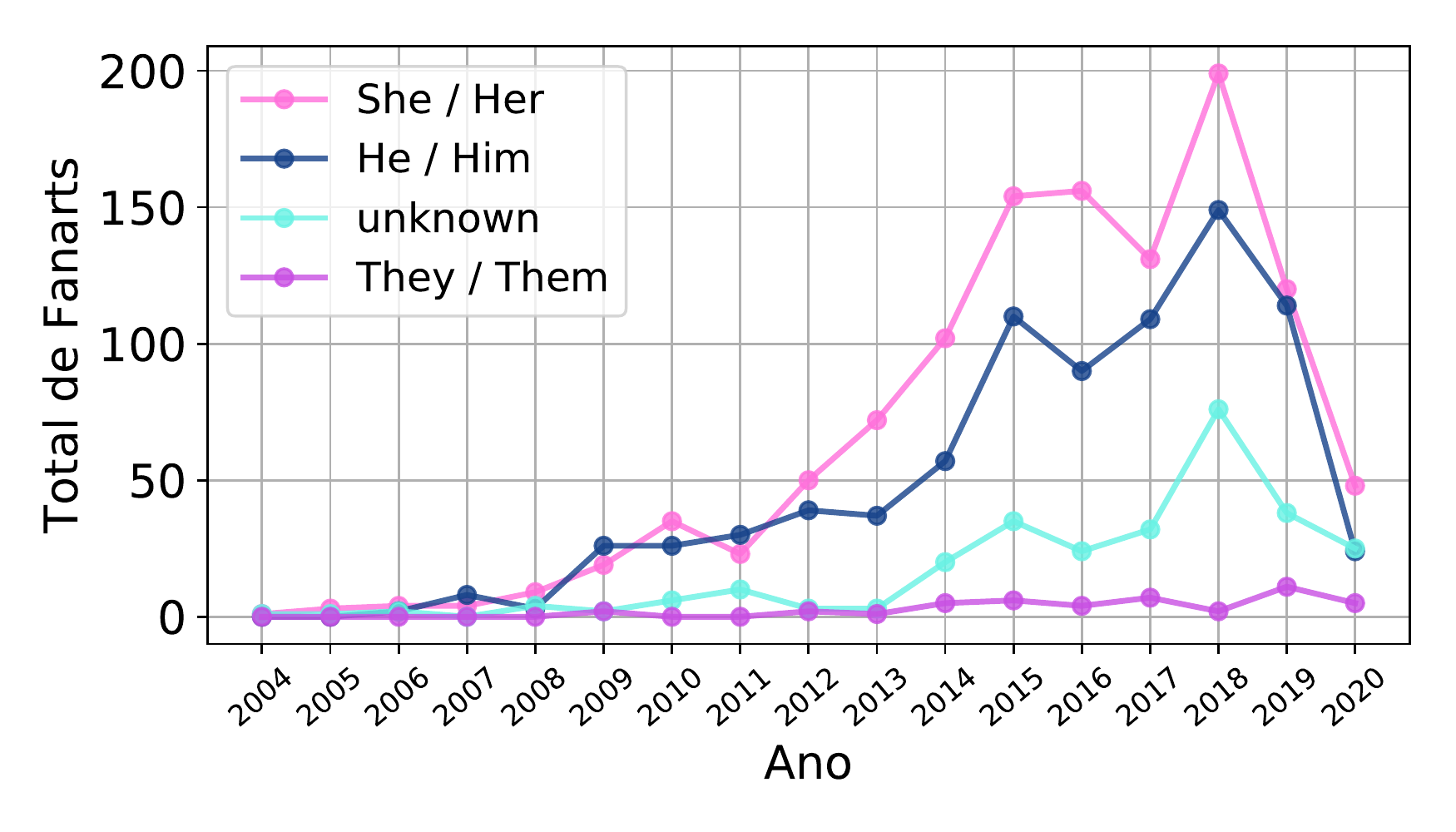}
        \vspace{-0.4cm}
         \caption{\textit{Fanarts} por ano de acordo com o gênero dos autores.}
        \vspace{-0.1cm}
        \label{fig:fanart_year}
        \vspace{-0.3cm}
\end{figure}

Na Fig.~\ref{fig:fanart_year} revisitamos a quantidade de \textit{fanarts} coletadas no dA por ano, porém aqui fizemos uma divisão conforme o gênero informado pelo artista. É interessante observamos que entre 2004 e 2012 não havia muita diferença entre a quantidade de publicações entre homens e mulheres, mas a partir do ano do lançamento da expansão \textit{Mists of Pandaria} (2012) e do livro Marés de Guerra (Christie Golden, 2012), protagonizados por Jaina, houve um aumento considerável da produção feminina, com maioria considerável entre 2015 e 2018\footnote{Observe que a coleta foi realizada em Junho de 2020, portanto os valores no gráfico não representam o total postado naquele ano.}. Entretanto, vemos que a produção de artistas ``They/Them'' não acompanhou esse crescimento, ficando sempre com uma baixa presença na comunidade. 

\begin{figure*}[t!]
    \centering
    \begin{minipage}[h!]{.30\linewidth}
        \includegraphics[width=0.99\linewidth]{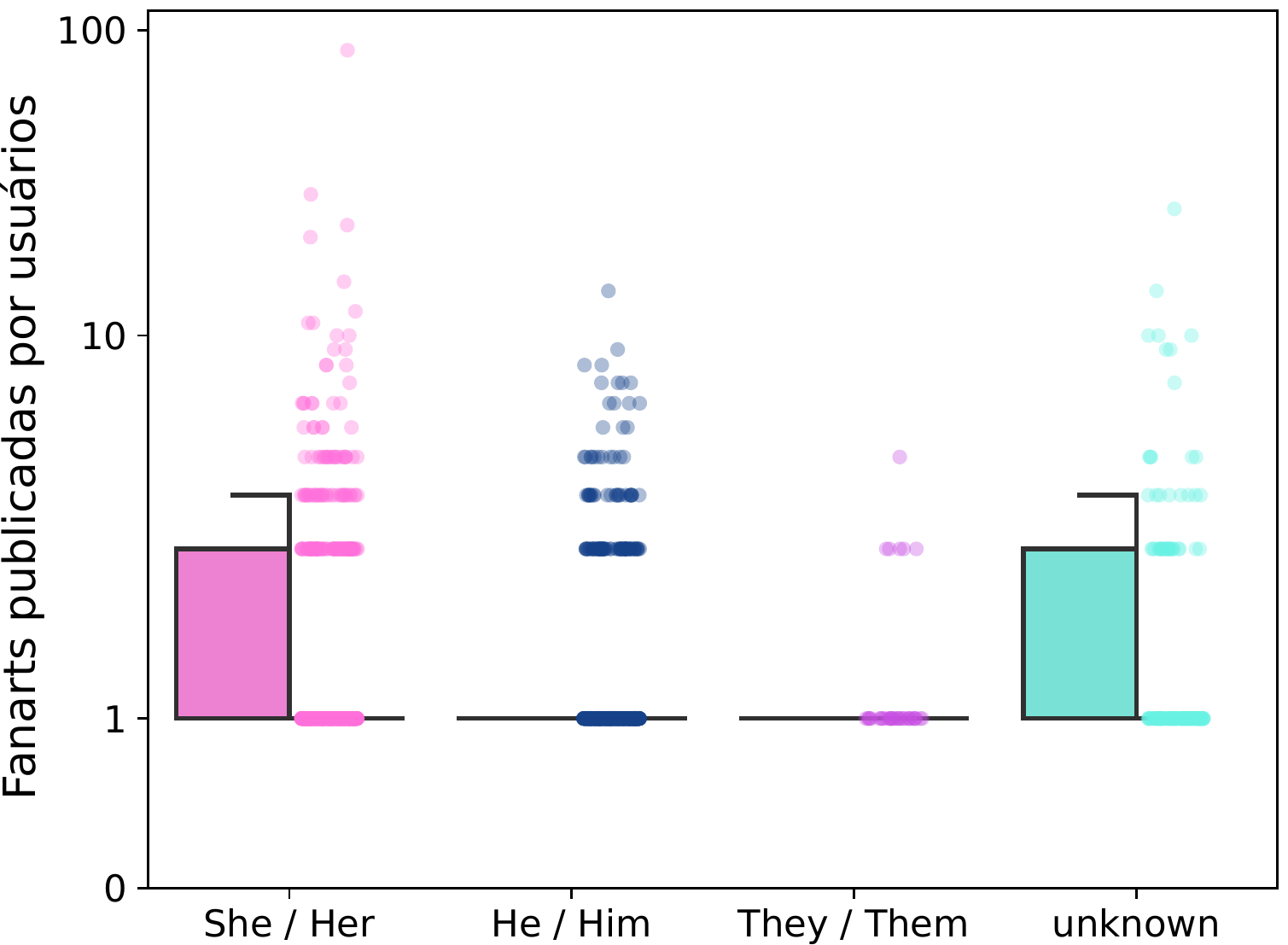}
    \caption{Boxplot com  quantidade de \textit{fanarts} de Jaina postada por artista.}\label{fig:cdf_artistas}
    \end{minipage}
    \hfill
    \begin{minipage}[h!]{.31\linewidth}
        \includegraphics[width=1.00\linewidth]{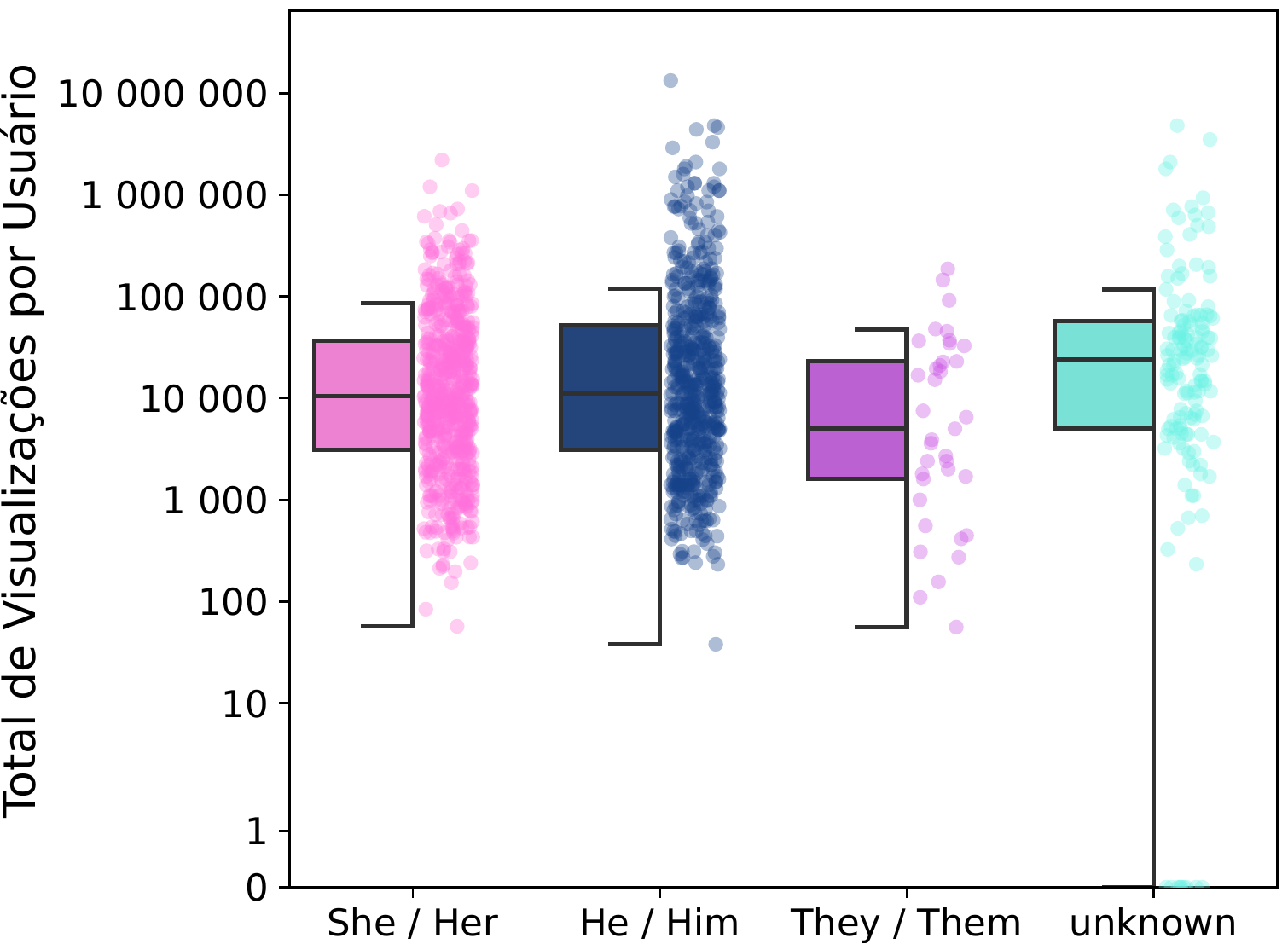}
        \caption{Boxplot com total de visualizações por artista.}\label{fig:cdf_views}
    \end{minipage}
    \hfill
    \begin{minipage}[h!]{.30\linewidth}
        \includegraphics[width=1.00\linewidth]{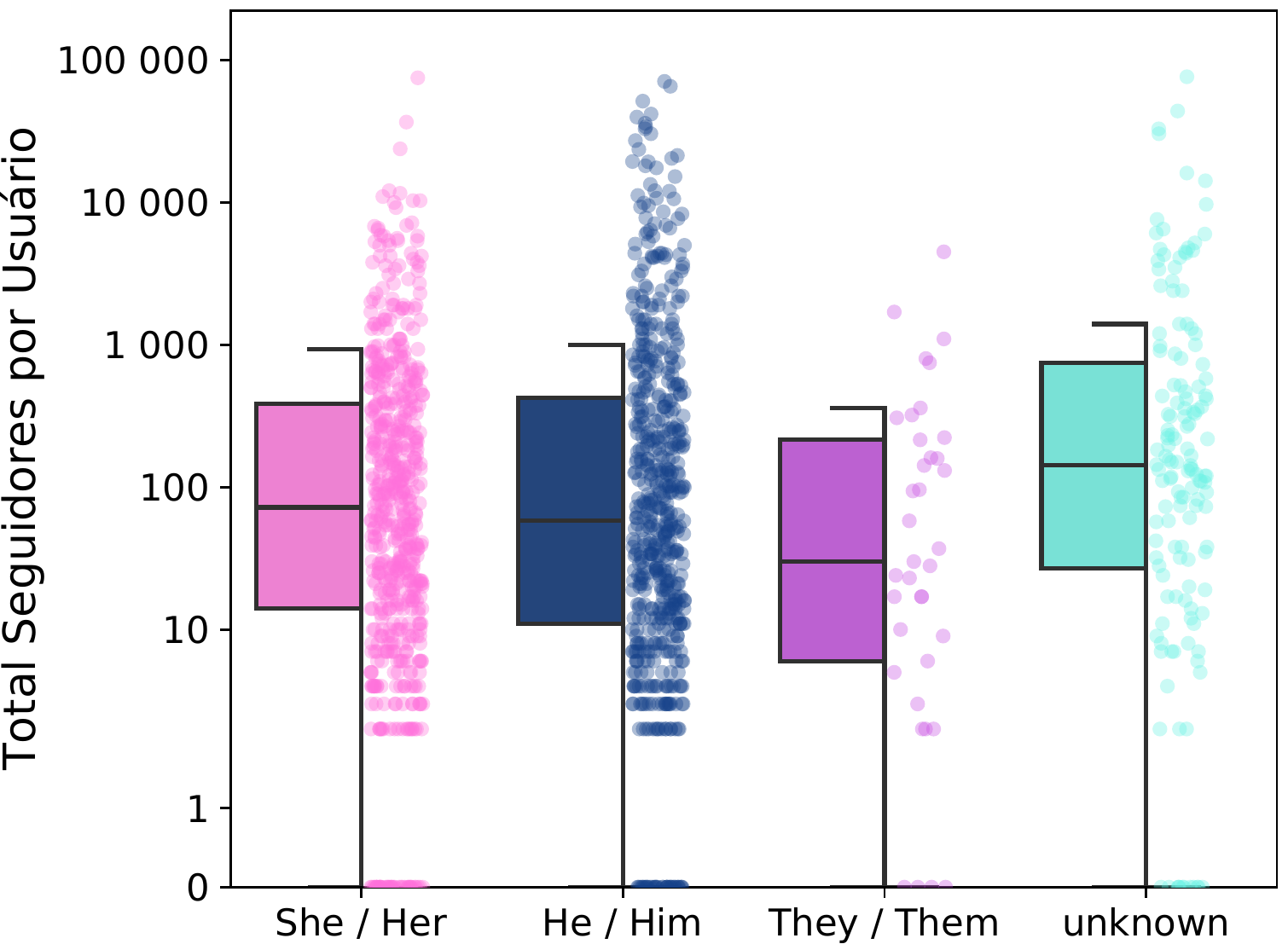}
        \caption{Boxplot com total de seguidores por artista.}\label{fig:cdf_watchers}
    \end{minipage}
    \vspace{-0.5cm}
\end{figure*}

Como existe uma abundante quantidade de publicações e cada \textit{fanart} possui seus dados individuais, foi preciso pensar uma forma clara e sucinta de apresentar todos os dados para analisá-los. Para enxergarmos as relações de produção e popularidade na plataforma, foi utilizado um boxplot para visualizarmos as distribuições dos dados. Nele, podemos ver, ordenada por gênero, a distribuição da quantidade de artes, seguidores e visualizações do nosso corpus de artistas, através do diagrama com mínimo, máximo e medianas para cada atributo.
Na Fig.~\ref{fig:cdf_artistas}, é apresentada a quantidade de \textit{fanarts} produzidas por usuário. Observamos que pelo menos 50\% das pessoas de cada gênero postou apenas uma única \textit{fanart} no conjunto analisado. Usuários que se identificam com ``\textit{They/Them}'' foram os que menos postaram artes, devido a constituírem também o grupo com menor representação na rede. Aqui vemos que usuários femininos tem uma tendência a postarem mais artes, com um pouco mais da metade das artistas possuindo de duas a 10 artes sobre Jaina na coleção.

Ao investigarmos estas artistas, percebemos que são em grande parte mulheres fizeram \textit{cosplay} e publicaram na plataforma. Elas costumam publicar séries de fotos vestidas como a personagem, o que resulta em um número maior de obras destes perfis em comparação aos outros que postam artes. Essas imagens, que verificamos manualmente, são também de cenas teatrais, não apenas registros fotográficos de confecção de \textit{cosplays}. Isso pode apontar para uma grande identificação das mulheres com a personagem, a ponto de se fantasiarem e produzirem uma série enquadramentos cênicos distintos performando como Jaina. 

Outras distribuições analisadas são os números de visualizações e de seguidores por usuário, apresentados respectivamente nas Fig.~\ref{fig:cdf_views} e Fig.~\ref{fig:cdf_watchers}. Esses dados ajudam a entender um pouco a diferença de popularidade entre os usuários nessa rede. Vemos que enquanto alguns usuários têm poucos seguidores e visualizações, outros possuem quase 100 mil seguidores e mais de um milhão de visualizações de seus trabalhos.
Os usuários identificados pelo pronome ``\textit{They/Them}'' ficam atrás tanto em número de seguidores quanto visualizações. Enquanto somente 10\% dos artistas mulheres e homens possuem menos de 1000 visualizações, aproximadamente 25\% dos artistas ``They/Them'' possuem essa baixa quantidade de visualizações. Ainda notamos que nenhum deles tem mais de um milhão de visualizações, enquanto temos 22 perfis masculinos com mais de 1 milhão de \textit{views} (incluindo o perfil com mais visualizações com 13.4 milhões). Quatro perfis \textit{unknown} e três perfis ``\textit{She/Her}'' também ultrapassaram a marca de um milhão de visualizações.

Vemos ainda nas distribuições de boxplot que o limite superior da distribuição dos homens é levemente mais elevado em comparação aos demais, tanto em visualizações como em seguidores. Isso demonstra que homens tem uma tendência de possuir perfis mais populares na plataforma, conquistando maior número de seguidores e visualizações. Além disso, \textit{fanarts} não são uma forma de mídia exclusiva do usuário desconhecido com poucas visualizações, sendo também o formato de produção escolhido por artistas já renomados. Todos os sujeitos ocupam um espaço no fandom, mas não podemos desconsiderar que alguns usuários podem ter mais peso nas disputas narrativas sobre a personagem, visto que o perfil possui uma visibilidade muito maior na plataforma.

\begin{figure}[t]
    \centering 
    \includegraphics[width=0.79\linewidth]{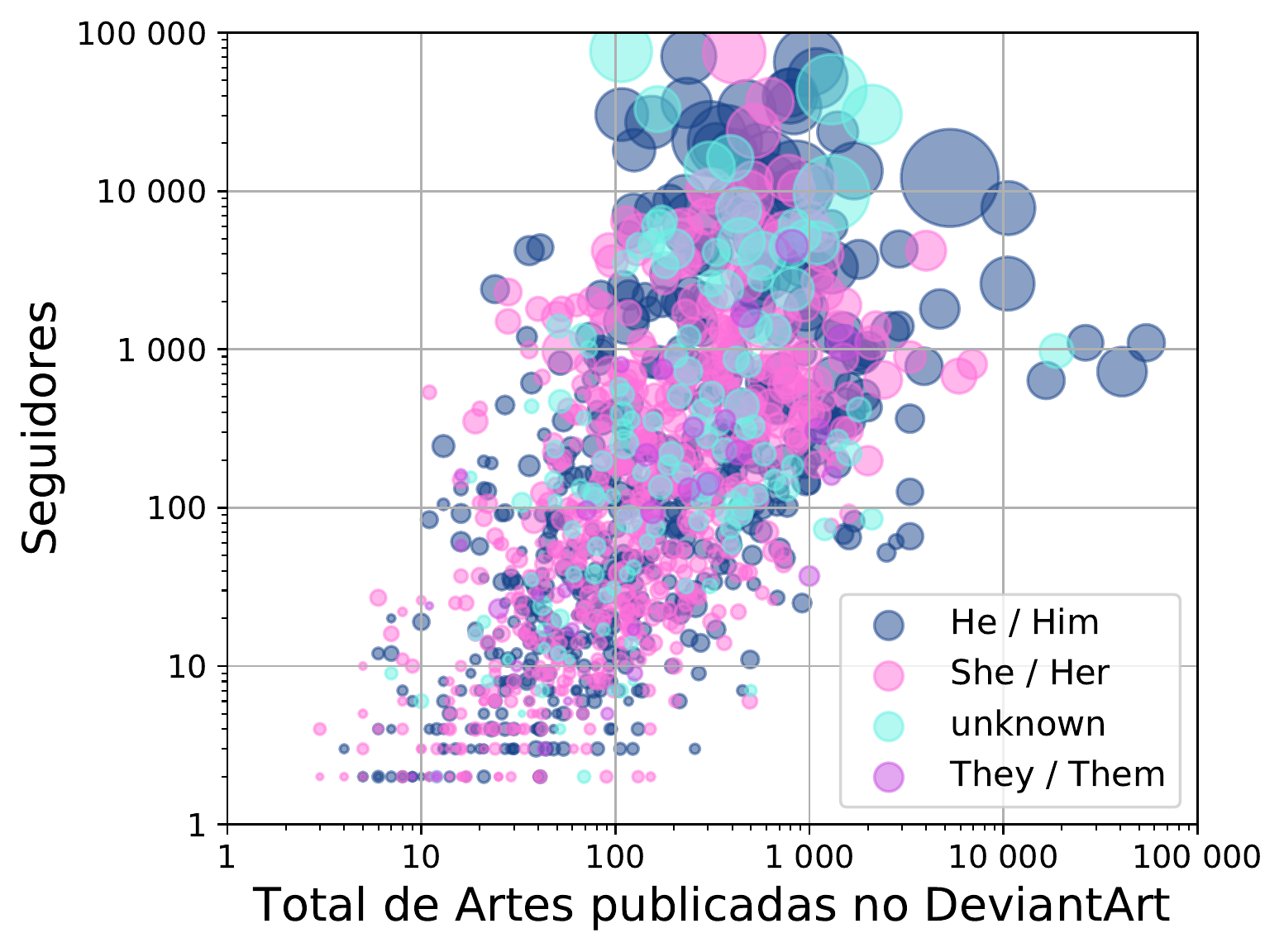}
        \vspace{-0.1cm}
    \caption{Relação entre número de trabalhos publicados e seguidores dos artistas que publicaram \textit{fanarts} de Jaina no DeviantArt.}
        \vspace{-0.7cm}		
    \label{fig:scatter}
\end{figure}

Para enxergarmos um pouco melhor como se relacionam esses dados de popularidade e a discrepância de gênero dos usuários, apresentamos a Fig.~\ref{fig:scatter}, que mostra a relação entre número de seguidores, visualizações, e total de trabalhos publicados pela conta (não apenas \textit{fanarts}). Cada círculo representa um usuário, quanto mais para direita, mais artes o usuário possui na plataforma, e quando mais para cima, maior o número de seguidores. O tamanho da bolinha corresponde à quantidade de visualizações.

À medida que nos aproximamos dos eixos de maior visualização e seguidores, há uma maior concentração de círculos azuis no gráfico (próximos ao topo e à direita), o que evidencia que usuários masculinos possuem maior popularidade, com mais artes publicadas e seguidores geralmente. Vemos aqui também que usuários com poucos trabalhos e/ou poucos seguidores dificilmente alcançam muitas visualizações.
Assim, podemos inferir que na disputa narrativa do fandom, as influências dos usuários tem pesos diferentes devido à dinâmica da plataforma. Artistas mais famosos, mais reconhecidos, com mais seguidores, podem acabar ditando o caminho do fandom, enquanto visões geradas por aqueles desconhecidos acabam ficando esquecidas dentro da plataforma.
Considerando as diferenças encontradas em relação ao gênero dos autores, essa disputa tem um significado ainda maior. Apesar da quantidade de usuários femininos ser ligeiramente maior do que masculinos, e mulheres postarem mais artes no geral, são as \textit{fanarts} de perfis masculinos que alcançam maior visibilidade na rede. Usuários ``They/Them'' ficam ainda mais atrás nesta disputa dentro do dA, mostrando mais uma vez como o gênero impacta na recepção da produção dos fãs e é engendrado pela plataforma.

\begin{figure}[t!]
    \centering
    \begin{minipage}[h!]{.48\linewidth}
        \includegraphics[width=0.85\linewidth]{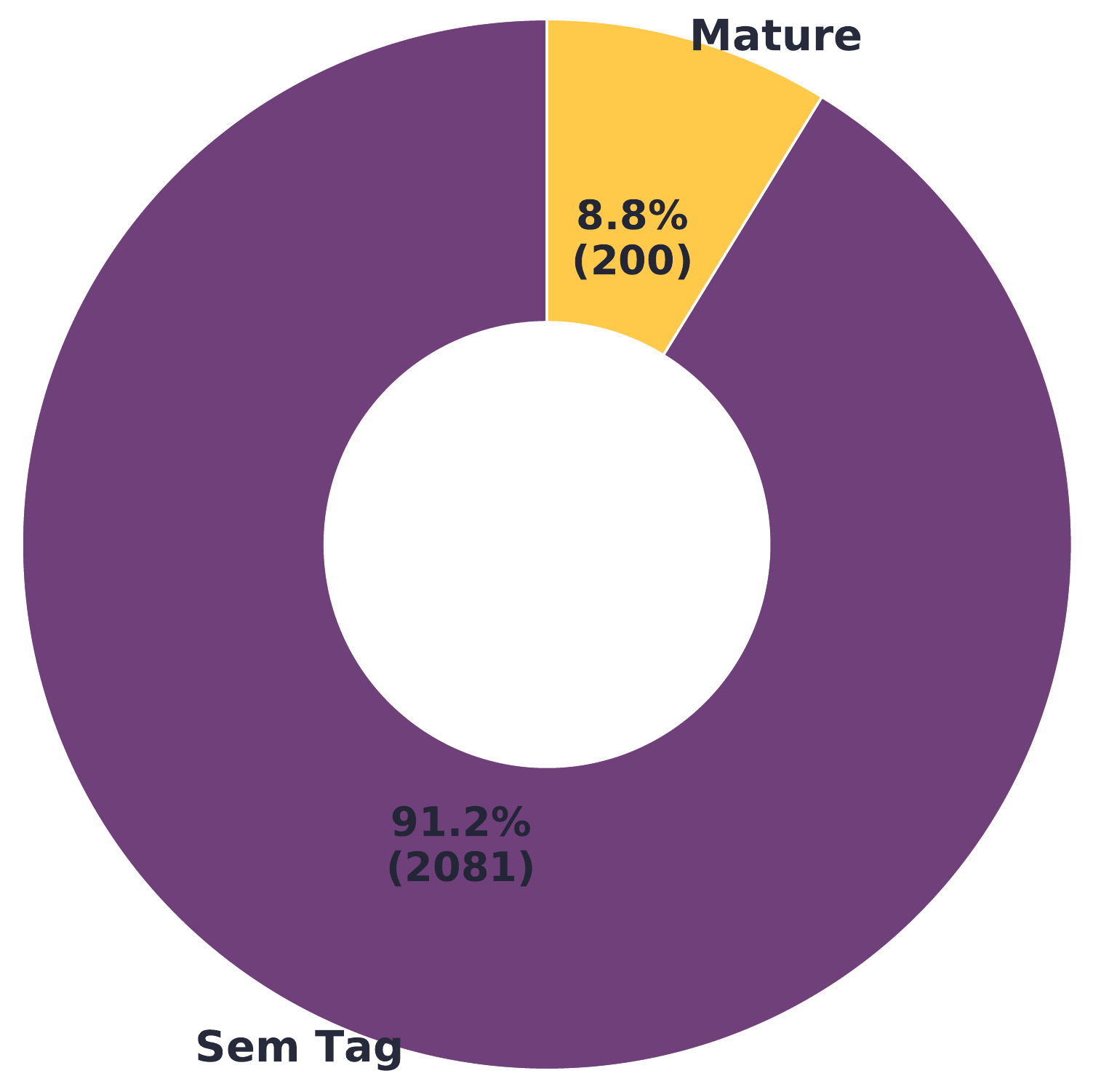}
        \caption{Quantidade de \textit{fanarts} de Jaina rotuladas como ``Mature''.}
        \label{fig:nsfw_pie}
    \end{minipage}
    \hfill
    \begin{minipage}[h!]{.49\linewidth}
        \includegraphics[width=1.0\linewidth]{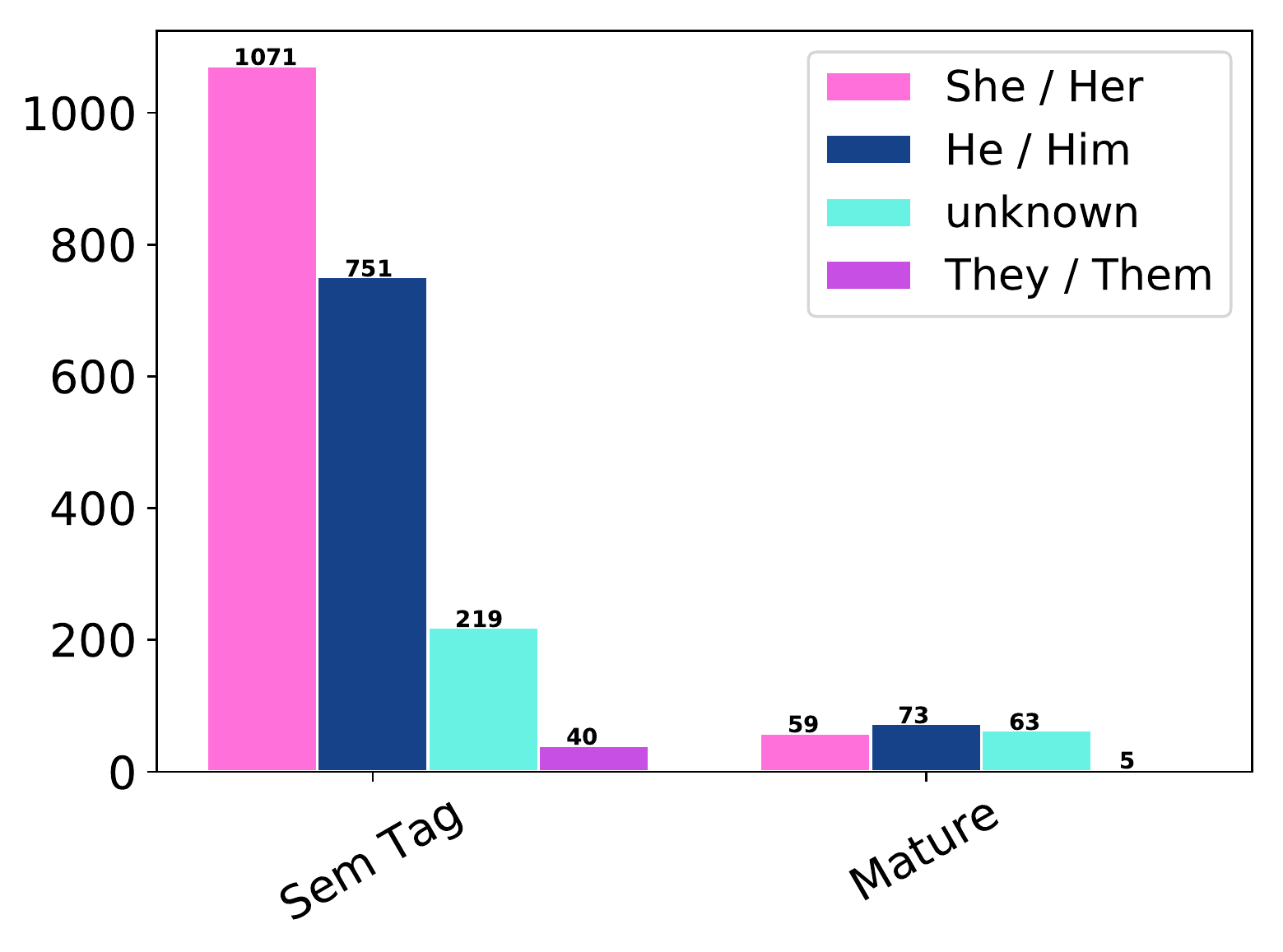}
        \caption{Divisão de gênero dos autores entre as \textit{fanarts} ``Mature''.}
        \label{fig:nsfw_gender}
    \end{minipage}
    \vspace{-0.5cm}
\end{figure}

Quando um artista posta uma obra na plataforma do dA, ele pode marcar o conteúdo como ``\textit{Mature}''. Esse tipo de conteúdo é direcionado a um público adulto e o usuário precisa habilitar a exibição para vê-las. Porém, mesmo existindo um guia de como realizar a categorização, fica a cargo de cada usuário determinar o que é ou não conteúdo adulto em seu perfil. Na Fig.~\ref{fig:nsfw_pie}, vemos que 8.8\% das \textit{fanarts} foram marcadas como conteúdo adulto, o que representa 200 publicações. Em Fig.~\ref{fig:nsfw_gender}, exibimos a relação de gênero do artista com a categoria de arte adulta. Embora existam mais autoras mulheres, quando olhamos apenas os trabalhos adultos, a maioria dos artistas se identificam com gênero masculino, seguido por usuários ``\textit{unknown}''. Já dentre os trabalhos sem restrição, é evidente a maioria de artistas mulheres, quase cinco vezes maior do que de gênero desconhecido.

Através de uma investigação manual das artes ``\textit{Mature}'' mais populares, constatamos que a maioria são imagens relacionadas à venda, com menções na descrição e/ou título a plataformas de comercialização de arte, como o Patreon, Gumroad ou Ko-fi\footnote{Disponíveis respectivamente em: \url{https://www.patreon.com/},\\ \url{https://gumroad.com/}, e \url{https://ko-fi.com/}}.
%
Por exemplo, a \textit{fanart} adulta mais popular, com maior número de favoritos, mostra Jaina junto à personagem Sylvana\footnote{Jaina Proudmoore x Sylvanas Windrunner -- \url{https://www.deviantart.com/flowerxl/art/Jaina-Proudmoore-x-Sylvanas-Windrunner-734365067}}. Em sua descrição, na página do artista, é informado que a \textit{fanart} foi feita sob encomenda para outra pessoa, com outras informações sobre vendas e amostras de artes. A \textit{fanart} enquadra as duas personagens se beijando, semi-nuas, com ênfase no tema sexual; elas são representadas sobre um fundo sem cenário (mas com uma luz holofote), os corpos estão em evidência, os seios e bundas são desenhados com realce (recebendo, inclusive, iluminação especular) e, além disso, foram desenhadas para venda. Essa imagem-espetáculo transforma o relacionamento lésbico das duas personagens em um fetiche submisso a um olhar masculino, e a venda revela uma dinâmica de comoditização do corpo feminino, ligada ao sistema capitalista.



\begin{figure}[t]
    \centering 
    \includegraphics[trim={0.1cm 5.5cm 0.1cm 5cm},clip, width=1.00\linewidth]{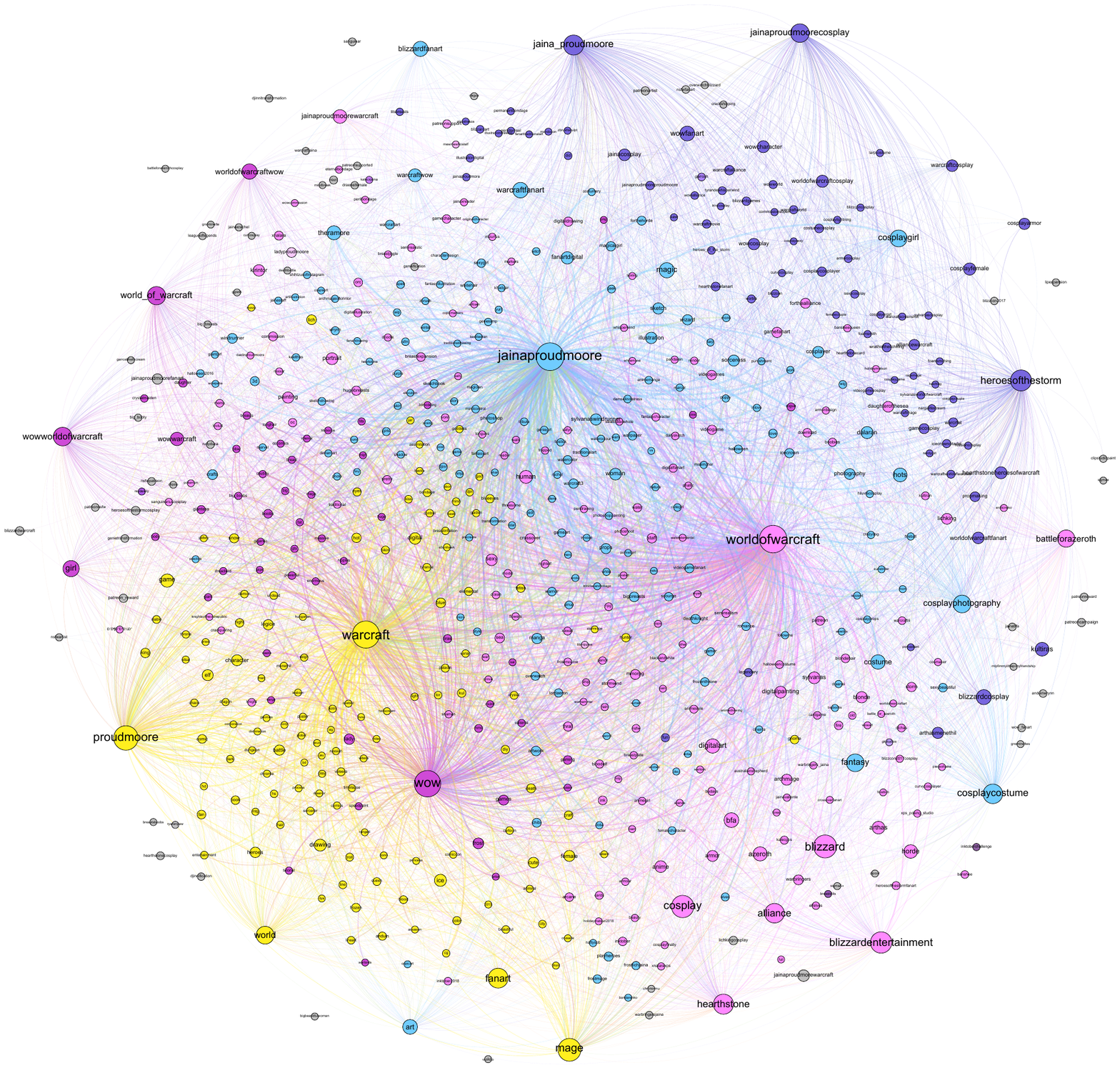}
    \caption{Rede de Tags das \textit{fanarts} de Jaina Proudmoore publicadas no Deviantart.}
        \vspace{-0.35cm}
    \label{fig:grafo_deviantart}
        \vspace{-0.4cm}
\end{figure}

Cada \textit{fanart} no dA conta com uma lista de tags e, para visualizarmos melhor esse conjunto de etiquetas atribuídas a cada obra, foi possível reconstruir a rede de tags desses trabalhos e enxergar como elas se relacionam, de forma que cada uma representa um ponto num grafo. Dois desses pontos conectados significa que existe algum trabalho que tem ambas as tags simultaneamente. Essa rede nos permite tanto visualizar as etiquetas mais evidentes no conjunto de dados, quanto suas interações com outras. A Fig.~\ref{fig:grafo_deviantart} exibe o grafo construído com auxílio da ferramenta \textit{Gephi}. 
%
Em comparação às tags das \textit{fanfics}, que possuiam etiquetas mais populares muito atreladas ao conteúdo da obra (\textit{angst}, \textit{fluff}, etc), no dA temos as tags mais populares um pouco mais genéricas, como \textit{WorldOfWacraft} e \textit{Fanart}, o que sugere uma categorização mais ampla para indexação do trabalho ao invés de descrição do conteúdo. Ainda nesse sentido, é interessante também notar a presença de tags sobre o processo de criação das artes, tais como \textit{painting} e \textit{zbrush}, que se referem ao estilo ou ferramenta utilizada para produção. Essas tags mais relacionadas com a feitura do trabalho, junto de outras como \textit{patreonnsfw}, \textit{patreon reward}, \textit{patreoncampaign}, \textit{commissionsopen} aproximam a \textit{fanart} de um produto comercial.

As tags usadas para descrever o conteúdo das \textit{fanarts} falam sobre os corpos das personagens, com ênfase em partes associadas à sexualidade --  \textit{hugebreasts}, \textit{boobs}, \textit{big boobs}, \textit{bikini}, \textit{hot}, \textit{tits}, \textit{bondage}, \textit{erotic}, \textit{fetish} --, o que demonstra um interesse por parte dos artistas em desenharem imagens com esse teor e utilizarem o desejo por representações sexuais para popularizar seus trabalhos. Aqui é interessante observar algumas outras etiquetas que aparecem próximas a essas, como \textit{breastinflation}, \textit{enormous}, \textit{giantess}, \textit{gigantic}, \textit{chubby}, relacionadas a fetiches sexuais. Essas tags diferem das etiquetas sobre conteúdo sexual do AO3, pois têm maior presença de elementos corporais e de fetiches específicos.



\section{Considerações Finais}
\vspace{-0.1cm}
Ao analisarmos o fandom de World of Warcraft através de plataformas de publicação de \textit{fanfics} e \textit{fanarts} e recuperarmos alguns dos estudos acerca do jogo, é possível notarmos que a experiência dos usuários e os modos como interagem com WoW estão fortemente engendrados pela transmidialidade da franquia e pelas plataformas digitais de publicação de conteúdo, não somente aqueles oficiais, mas também pelos gerados por fãs. 
O Archive of Our Own é uma plataforma criada com o intuito de ser um repositório de livre acesso a autores e leitores, diferente do DeviantArt, que possui recursos para comercialização das artes. Isso implica diretamente em como essas redes organizam os enunciados produzidos pelos fãs acerca dos jogos, em particular no caso do dA, que acentua a presença masculina no fandom, apesar da maioria feminina. 
As produções no dA feitas por homens são geralmente permeadas pela lógica de venda, contrariando a percepção comum de que fãs produzem seguindo uma economia de presente. Percebemos que os usuários masculinos tem maior tendencia a adequar suas produções ao mercado e buscar maneiras de populariza-las, utilizando vários termos para indexação e elementos de hipersexualização capazes de atrair público pagante. Essas obras com viés mercadológico são absorvidas e disseminadas com facilidade pela plataforma e conectam-se a outros sites (como Patreon e Gumroad). Em contrapartida, as obras femininas são fortemente afetivas, muitas com os próprios corpos das autoras em cena. Elas também são as maiores produtoras de \textit{fanarts}. Entretanto, não alcançam a mesma popularidade que os homens. Usuários ``\textit{They/Them}'', que podemos entender como pessoas \textit{queer}, ficam ainda mais ocultos no dA.  

Apesar da forte cultura masculina e heteronormativa que permeia Warcraft, no AO3 a maioria dos trabalhos retrata relacionamentos sexuais e narrativas dissidentes do cânone. São obras afetivas que demonstram haver um público LGBTQI+ no fandom que se interessa pelo jogo e busca explorar sexualidades e identidades de gênero diversas através dele. Diferente dos trabalhos do dA, esses temas emergem no AO3 com mais facilidade, fazendo crer que os fãs enxergam a plataforma como um espaço seguro para tratarem de suas identidades.
No AO3 parece haver o predomínio de uma produtividade semiótica e enunciativa, em que há um grande investimento afetivo dos autores em suas obras e a postagem e compartilhamento entre pares na plataforma. Já no caso do dA, é possível notarmos uma produtividade textual, de modo que os trabalhos são compartilhados entre plataformas e comercializados. 

Mulheres e outros públicos se engajam em grande volume com o jogo, participando ativamente da história ao criar narrativas e artes. O desenvolvimento de enredos cânones com maior participação feminina surge como um incentivo à criação, como ocorreu no caso de Jaina e Sylvanas. A representação feminina importa, seja para abrir horizontes de possibilidades sobre relacionamentos fora da heteronomartividade, seja em termos de identificação das mulheres, como no caso das \textit{cosplayers} que se fantasiam de Jaina Proudmoore. Ainda assim, esses sujeitos são invisibilizados pela lógica capitalista das plataformas.

\balance
\printbibliography

\end{document}